\documentclass[aps,prb,showpacs,twocolumn,preprintnumbers,amsmath,amssymb,showkeys]{revtex4} 

\usepackage{graphicx}
\usepackage{dcolumn}
\usepackage{bm}
\begin{document}

\title{\emph{Poly}-MTO, \{(CH$_{3}$)$_{0.92}$ReO$_{3}$\}$_{\infty}$, a Conducting Two-Dimensional Organometallic Oxide}

\author{R. Miller}
\author{E.--W. Scheidt}
\author{G. Eickerling}
\author{C. Helbig}
\author{F. Mayr}
\author{R.~Herrmann}
\author{W. Scherer}

\email{Wolfgang.Scherer@Physik.Uni-Augsburg.de}

\affiliation{Chemische Physik und Materialwissenschaften,
Universit\"{a}t Augsburg, 86159 Augsburg, Germany}

\author{H.--A. Krug von Nidda}

\affiliation{Experimentalphysik V, Elektronische Korrelationen und
Magnetismus, Universit\"{a}t Augsburg, 86159 Augsburg, Germany}

\author{V. Eyert}
\author{P. Schwab}

\affiliation{Theoretische Physik II, Universit\"{a}t Augsburg,
86159 Augsburg, Germany}

\date{\today}

\begin{abstract}
Polymeric methyltrioxorhenium,
\{(CH$_{3}$)$_{0.92}$ReO$_{3}$\}$_{\infty}$ (\emph{poly}-MTO), is
the first member of a new class of organometallic hybrids which
adopts the structural pattern and physical properties of classical
perovskites in two dimensions (2D). We demonstrate how the
electronic structure of \emph{poly}-MTO can be tailored by
intercalation of organic donor molecules, such as
tetrathiafulvalene (TTF) or bis-(ethylendithio)-tetrathiafulvalene
(BEDT-TTF), and by the inorganic acceptor SbF$_3$. Integration of
donor molecules leads to a more insulating behavior of
\emph{poly}-MTO, whereas SbF$_3$ insertion does not cause any
significant change in the resistivity. In particular, with
increasing donor intercalation the metallic behavior of the parent
compound, \emph{poly}-MTO, becomes surprisingly suppressed leading
to an insulator at TTF (BEDT--TTF) donor concentrations above
$50\%$ ($25\%$). The resistivity data of pure \emph{poly}-MTO
exhibit a crossover from metallic (d$\rho$/d$T
>$ 0) to insulating (d$\rho$/d$T <$ 0) behavior at a
characteristic temperature around $T_{\textrm{min}}\simeq 38$\,K.
Above $T_{\textrm{min}}$ the resistivity $\rho(T)$ is remarkably
well described by a two-dimensional electron system. Below
$T_{\textrm{min}}$ an unusual resistivity behavior, similar to
that found in doped cuprates, is observed: The resistivity
initially increases approximately as $\rho \sim$ ln$(1/T$) before
it changes into a $\sqrt{T}$ dependence below 2\,K. As an
explanation we suggest a crossover from purely two-dimensional
charge-carrier diffusion within the  \{ReO$_2$\}$_{\infty}$ planes
at high temperatures to three-dimensional diffusion at low
temperatures in a disorder-enhanced electron-electron interaction
scenario (Altshuler-Aronov correction). Furthermore, a linear
positive magnetoresistance was found in the insulating regime,
which is caused by spatial localization of itinerant electrons at
some of the Re atoms, which formally adopt a $5d^1$ electronic
configuration. X-ray diffraction, IR- and ESR-studies, temperature
dependent magnetization and specific heat measurements in various
magnetic fields suggest that the electronic structure of
\emph{poly}-MTO can safely be approximated by a purely 2D
conductor which is labile towards spatial localization of
electrons under formation of Re ($d^1$) centers in the presence of
a magnetic field.
\end{abstract}

\maketitle

\section{\label{sec:level1}Introduction}

The metal oxide polymeric methyltrioxorhenium
\{(CH$_{3}$)$_{0.92}$ReO$_{3}$\}$_{\infty}$ (\emph{poly}-MTO) is
an unique representative of an inherent conducting organometallic
polymer. For hydrogen-doped \emph{poly}-MTO samples
\{H$_{0.5}$[(CH$_{3}$)$_{0.92}$ReO$_{3}$]\}$_{\infty}$ obtained by
polymerization in aqueous solution a moderately high electrical
\emph{dc}-resistivity of 6\,m$\Omega$\,cm at room temperature and
a Pauli-like high-temperature magnetic susceptibility of $70
\times 10^{-6}$\,emu/mol was found in  earlier
studies.\cite{Herrmann:92,Herrmann:95} The increase of the
susceptibility with decreasing temperature was ascribed to an
antiferromagnetic coupling mechanism and the conductivity was
explained by the presence of demethylated Re atoms (approx. $8\%$
of all Re atoms are lacking a methyl group) and acidic hydrogen
atoms which formally act as a source of itinerant electrons of the
transition-metal oxide lattice.\cite{Fischer:94} The results of
band-structure calculations employing extended H\"uckel theory
were interpreted by Genin \emph{et al.}\cite{Genin:95} such that
these demethylated Re atoms, which formally represent
Re$^{\mathrm{VI}}$($d^1$) sites, are effectively oxidized and
their valence electrons are transferred to the band system. Only a
minor part of these electrons (0.05$\%$ Re atoms \cite{Miller:05})
remains located at the metal sites which are in the following
treated as Re($d^1$) centers. These paramagnetic centers build up
a two-dimensional (2D), diluted metal-oxide spin system in a
metal-like matrix. First experiments to enhance the electronic
conductivity of \emph{poly}-MTO by employing the organic donor
species tetrathiafulvalene (TTF) lead to an amazing result: a
crossover from metallic to insulating behavior with increasing TTF
contribution.\cite{Miller:05}

In a recently published paper \cite{Scheidt:05} it is observed
that the resistivity of \emph{poly}-MTO at low temperatures and in
high magnetic fields within the  \{ReO$_2$\}$_{\infty}$ planes
resembles the behavior of $\rho$ of the \{CuO$_2$\}$_{\infty}$
planes in Zn-doped high-$T_\mathrm{c}$ superconductors, e.g.
YBa$_{2}$Cu${}_3$O$_{7-\delta}$ (Ref.~\onlinecite{Segawa:99}) and
La$_{1.85}$Sr$_{0.15}$CuO$_{4}$ (Ref.~\onlinecite{Karpinska:00}).
At low temperatures a logarithmic divergence of $\rho \sim$
ln$(1/T$) over an extended temperature range is observed, followed
by a square root behavior ($\rho \sim \sqrt{T}$). In this
insulating regime, a positive, increasing magnetoresistance (MR)
is found. In the case of Zn-doped high-$T_\mathrm{c}$
superconductors the scattering centers are established by
nonmagnetic Zn atoms within the antiferromagnetic spin-correlated
\{CuO$_2$\}$_{\infty}$ planes, whereas the scattering centers of
\emph{poly}-MTO reflect the inverse situation: magnetic Re($d^1$)
centers are placed in nonmagnetic  \{ReO$_2$\}$_{\infty}$ planes.

Many high-$T_\mathrm{c}$ cuprates feature this logarithmic
divergence of the resistivity at low temperatures, when
superconductivity is entirely suppressed by e.g. applied magnetic
fields\cite{Segawa:99,Ando:95,Boebinger:96} or point defects, such
as Zn or Li impurities\cite{Segawa:99,Karpinska:00,Bobroff:99} or
oxygen and copper vacancies\cite{Rullier:01} (for a short
overview, see Ref.~\onlinecite{Luo:05}). Up to now various models
have been proposed to describe the ln($1/T$) behavior in the
insulating regime. Here we mention some of them: (i) In
Tl$_2$Ba$_2$CuO$_{\rm{6+\delta}}$ the defect scattering is purely
elastic and is accounted for by a 2D weak localization
theory.\cite{Rullier:01} (ii) A Kondo-like scattering is proposed
for underdoped YBa$_{2}$Cu${}_3$O$_{6.6}$ controlled by electron
irradiation or for the magnetic properties of
YBa$_{2}$Cu${}_3$O$_{6.7}$ induced by spinless defects like
$2.7\%$ Zn.\cite{Rullier:01} (iii) Varma suggested for single
layer Bi compounds a temperature-dependent impurity scattering
time in a marginal Fermi liquid.\cite{Varma:97} (iv) A
conventional electron-electron interaction in a 2D disordered
system\cite{Altshuler-Aronov:85} is also valid to describe the
electronic situation for different types of
cuprates.\cite{Ando:95,Li:02} However, there is still no consensus
about this issue. In this respect \emph{poly}-MTO may be a
promising model system to shed more light on the multifarious
discussion about the electron-scattering mechanism in cuprates.

In this article the exceptional 2D character of the structure as
well as of the charge-carrier transport of \emph{poly}-MTO is
consistently revealed by density-functional theory (DFT)
calculations, x-ray diffraction, IR-studies, specific heat, and
electrical resistivity measurements. For intercalated
\emph{poly}-MTO we present a phase diagram separating the metallic
phase (d$\rho$/d$T >$ 0) from the insulating one (d$\rho$/d$T <$
0), taking into account guest molecules like the organic donor
molecules tetrathiafulvalene (TTF)
and the inorganic acceptor compound SbF$_3$. Furthermore,
ESR-results support the presence of spatially localized electrons as
revealed by the presence of Re($d^1$) centers.\cite{Scheidt:05} In
addition, we present a detailed analysis of the temperature
dependence of the resistivity data in the insulating regime in
comparison to the Zn-doped high-$T_\mathrm{c}$ superconductor
La$_{1.85}$Sr$_{0.15}$CuO$_{4}$.\cite{Karpinska:00} In particular we
will show that the Altshuler-Aronov correction
\cite{Altshuler-Aronov:85} (disorder enhanced electron--electron
interaction)  works very well in describing the low-temperature
resistivity data for both, \emph{poly}-MTO and Zn-doped
La$_{1.85}$Sr$_{0.15}$CuO$_{4}$.

\section{\label{sec:level2}Preparation and Characterization}

\subsection{Synthesis}

As starting material for \emph{poly}-MTO we prepared
methyltrioxorhenium (MTO) in the following way:\cite{Herrmann:02}
Rhenium powder (10.43\,g, 56.0\,mmol) was suspended in 10\,ml of
water. Under cooling employing an ice-water bath and efficient
stirring, a total of 100\,ml of 35\% H$_2$O$_2$ was added in five
portions within two hours. After stirring for 1\,h at room
temperature, the mixture was heated to 80\,$^\circ$C for 2\,h and
then cooled to room temperature. A solution of AgNO$_3$ (10.0\,g,
58.8\,mmol) in 20\,ml of water was then added. After stirring for
30\,min, the precipitate was filtered off, washed with water (2
times 40\,ml) and diethylether (2 times 60\,ml), and dried for 5\,h
at room temperature and 0.2\,mbar. The yield of AgReO$_4$ was
17.9\,g (89\%). The dry AgReO$_4$ was dissolved in 180\,ml of dry
acetonitrile under nitrogen, and 13.7\,ml (108.4\,mmol) of
trimethylchlorosilane were added, followed by tetramethyltin
(7.5\,ml, 54.0\,mmol). Stirring under nitrogen was continued for
16\,h. The precipitate was then filtered off under nitrogen with a
G3 glass filter and washed with acetonitrile (60\,ml). The solvents
were removed from the combined filtrates at 0.2\,mbar at room
temperature, and the semi-solid residue was washed with hexane (2
times 60\,ml). The solid residue was purified by sublimation at
0.2\,mbar/55\,$^\circ$C (oil bath temperature). The sublimation was
repeated once to obtain pure material, showing no impurities in the
$^1$H-NMR spectrum, giving correct C, H, Re elemental analysis, and
having a melting point of 106\,$^\circ$C. The yield was 9.06\,g
(65\% rel. Re metal).

\begin{figure}[b]
\begin{center}
\includegraphics[width=0.44\textwidth]{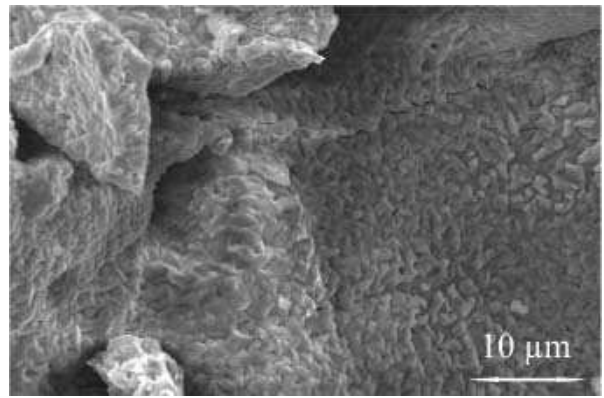}
\caption{Scanning electron microscope micrograph of
\emph{poly}-MTO microcrystalline grains.} \label{fig1}
\end{center}
\end{figure}

For the synthesis of \emph{poly}-MTO, two methods have been
established. The first consists in dissolving MTO
(CH$_{3}$ReO$_{3}$) in water under stirring at 80\,$^\circ$C for
two days.\cite{Herrmann:95_2} However, this is neither applicable
to the intercalation of TTF since this compound is not soluble in
water, nor to other reactive intercalates which often decompose in
the presence of water. A second method for \emph{poly}-MTO
synthesis has therefore been developed which is based on
auto-polymerization of MTO in the melt (melting point of pure MTO:
106\,$^\circ$C). The resulting polymer is a golden colored solid
compound. A scanning electron microscope image of \emph{poly}-MTO
is depicted in Fig.~\ref{fig1} indicating an assemble of
microcrystalline grains with typical diameters between 0.7\,$\mu$m
and 1.3\,$\mu$m which stick together.

The second technique allows for convenient and efficient
intercalation of a variety of organic, organometallic and inorganic
guest species. Thus, finely ground mixtures of MTO and the
intercalates (TTF, BEDT-TTF, SbF$_3$) in appropriate ratios are
heated in sealed ampoules at 120\,$^\circ$C or, more conveniently,
in closed screw cap glasses during two days (three days for the
samples with BEDT-TTF). The products were washed with an appropriate
solvent to remove unreacted starting materials (hexane for TTF, THF
for BEDT-TTF, and toluene for SbF$_3$), and dried under argon. The
samples were analyzed for their composition by elemental analysis
(C, H, S, using an Elementar Vario EL III apparatus), and ICP-OES
techniques (Re, S, Sb, with a Varian Vista MPX instrument). From the
analytical data, the formula and the molecular weight were
calculated as (CH$_3$)$_y$ReO$_3$(intercalate)$_x$, taking into
account that \emph{poly}-MTO shows a reduced content of methyl
groups ($y \leq 0.92$). The intercalated samples (formulated as
\emph{poly}-MTO + $x\%$\,TTF) prepared by this way form
bronze-colored solids for low TTF concentrations ($x < 40$) and
almost black powders for higher intercalation ratios. Whereas pure
\emph{poly}-MTO and its intercalated SbF$_3$ derivatives exist as
bronze-colored grains, the BEDT-TTF intercalated samples are
available solely as black powders.

\subsection{\label{DFT}Structural Characterization}

\begin{figure}[t]
\begin{center}
\includegraphics[width=0.44\textwidth]{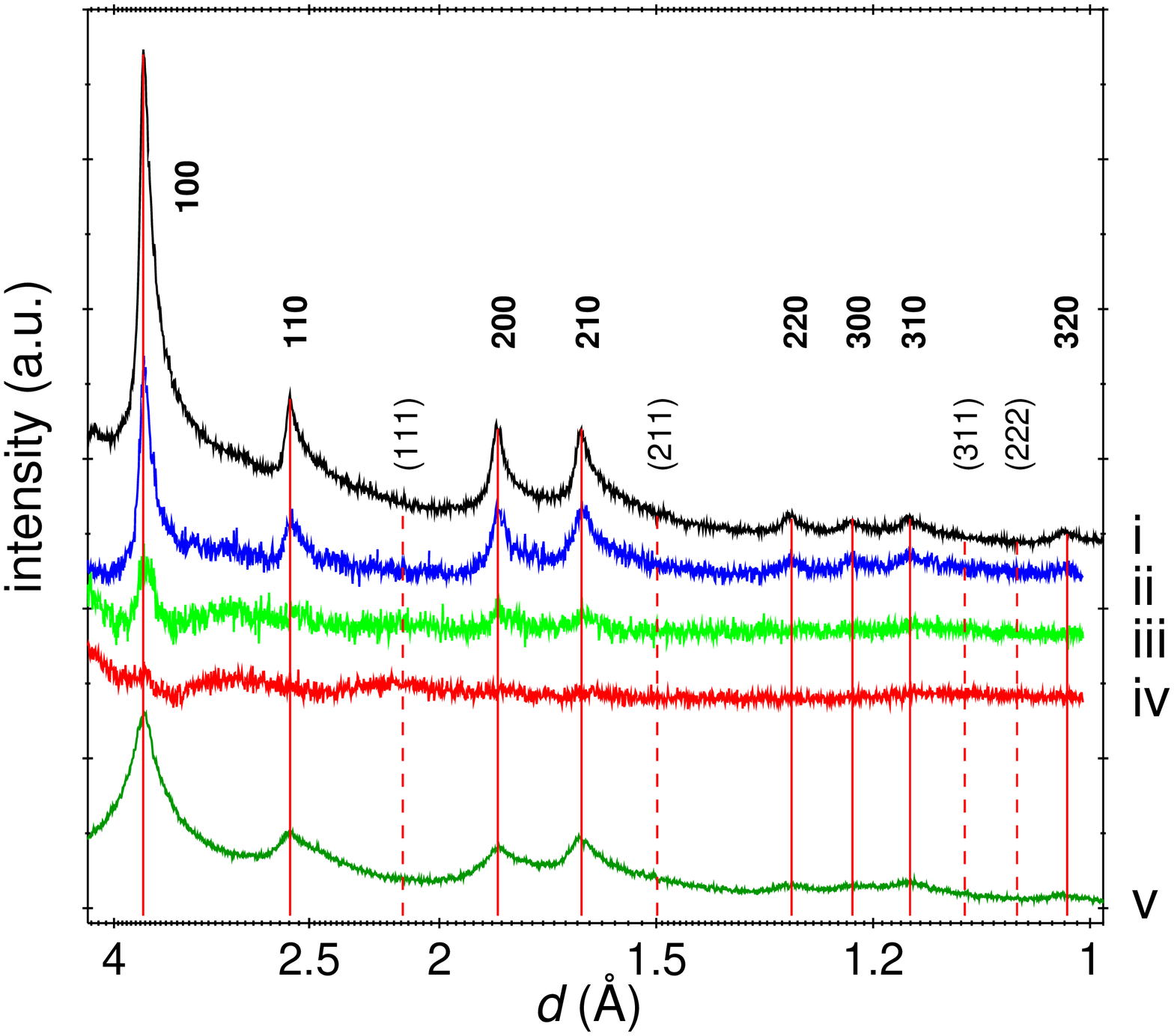}
\caption{X-ray powder diffraction patterns (Cu$\,K_{\alpha}$ and
Mo$\,K_{\alpha}$ radiation)\cite{radiation} of selected samples:
i) Parent compound \emph{poly}-MTO; ii) and iii) \emph{poly}-MTO
intercalated by 9\% and 29\% TTF, respectively; iv)
\emph{poly}-MTO intercalated by 25\% BEDT-TTF and v) by 1.1\%
SbF$_3$. The systematic absence of reflections \emph{hkl} with
$\emph{l} \neq 0$ indicates a layered structure without 3D
ordering for all samples.} \label{fig2}
\end{center}
\end{figure}

All samples have been characterized by x-ray powder diffractometry.
The diffraction pattern of these \emph{poly}-MTO samples exhibits
two remarkable features: First, all observed Bragg reflections of
the various samples can be indexed by cubic lattice parameters in a
narrow range of $3.66(5)$\,\AA $\leq a\leq 3.68(5)$\,\AA. These
lattice parameters are related but significantly different from
those reported for the cubic inorganic oxides ReO$_3$ ($a =
3.748$\,\AA)\cite{Herrmann:92,Meisel:32} and Re$_{1-x}$W$_x$O$_3$
($a = 3.7516(2)$\,\AA; $x = 0.25$).\cite{Helbig:05} We note further
the systematic absence of reflections \emph{hkl} for $l \neq 0$ (see
curve i in Fig.~\ref{fig2}, dashed lines). This is a strong evidence
for periodic ordering in these \emph{poly}-MTO samples to occur
exclusively in two dimensions.

A second hint for the presence of a layered structure without 3D
ordering is given by the asymmetry of the reflection profiles which
show a slow decay of Bragg intensity  in the direction of decreasing
$d$ values (cf. 100 and 110 reflection in Fig.~\ref{fig2}). This
peak shape asymmetry is a well-known indicator for layered compounds
displaying a turbostratic or 00\emph{l} defect stacking. The
reciprocal space construction for the resulting 2D diffraction
pattern consists of spread diffuse 00\emph{l} rods parallel to the
$c$ axis (stacking direction) which cause the smooth decay of the
Bragg intensities of each \emph{hk}0 reflection with increasing
diffraction angle $\theta$ or decreasing $d$-values.
\cite{Herrmann:95}

The observed diffraction pattern for \emph{poly}-MTO are therefore
in accord with the 2D space group \emph{p}4\emph{mm} and a square
unit cell. Hence, an idealized structural model for
\emph{poly}-MTO can be derived from the inorganic parent compound
ReO$_3$ (spacegroup \emph{Pm}$\bar{3}$\emph{m}) by adopting its
perovskite structure in two dimensions in form of
\{ReO$_{2}$\}$_{\infty}$ layers. Complementing the coordination
environment of the Re atom by one methyl (CH$_3$) and one oxo (O)
group leads to a layered network of CH$_3$ReO$_5$ octahedra
displaying an averaged Re--Re separation of about $3.67\pm
0.02$\,\AA.

Samples intercalated by the acceptor SbF$_3$ show no significant
change of the 2D character. Intercalation by TTF and BEDT-TTF
preserves the 2D structural character. However, the periodicity
within this structure becomes reduced with increasing donor
concentrations resulting in amorphous-like x-ray pattern (see, for
selected samples, curves ii--v in Fig.~\ref{fig2}).

\begin{figure}[t]
\begin{center}
\begin{minipage}[c]{0.19\textwidth}
\includegraphics[width=\textwidth]{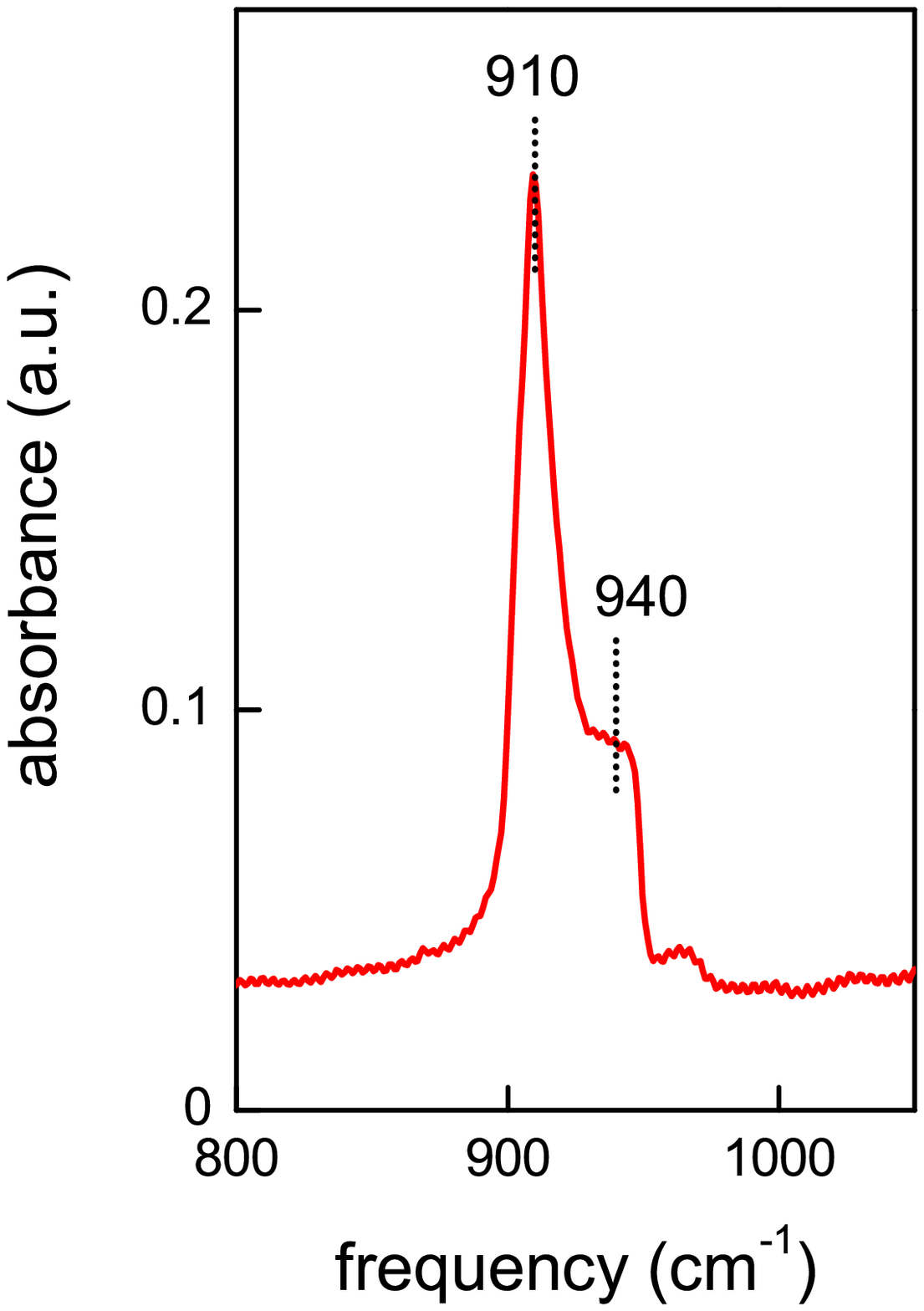}
\end{minipage}
\hfill
\begin{minipage}[c]{0.28\textwidth}
\includegraphics[width=\textwidth,clip]{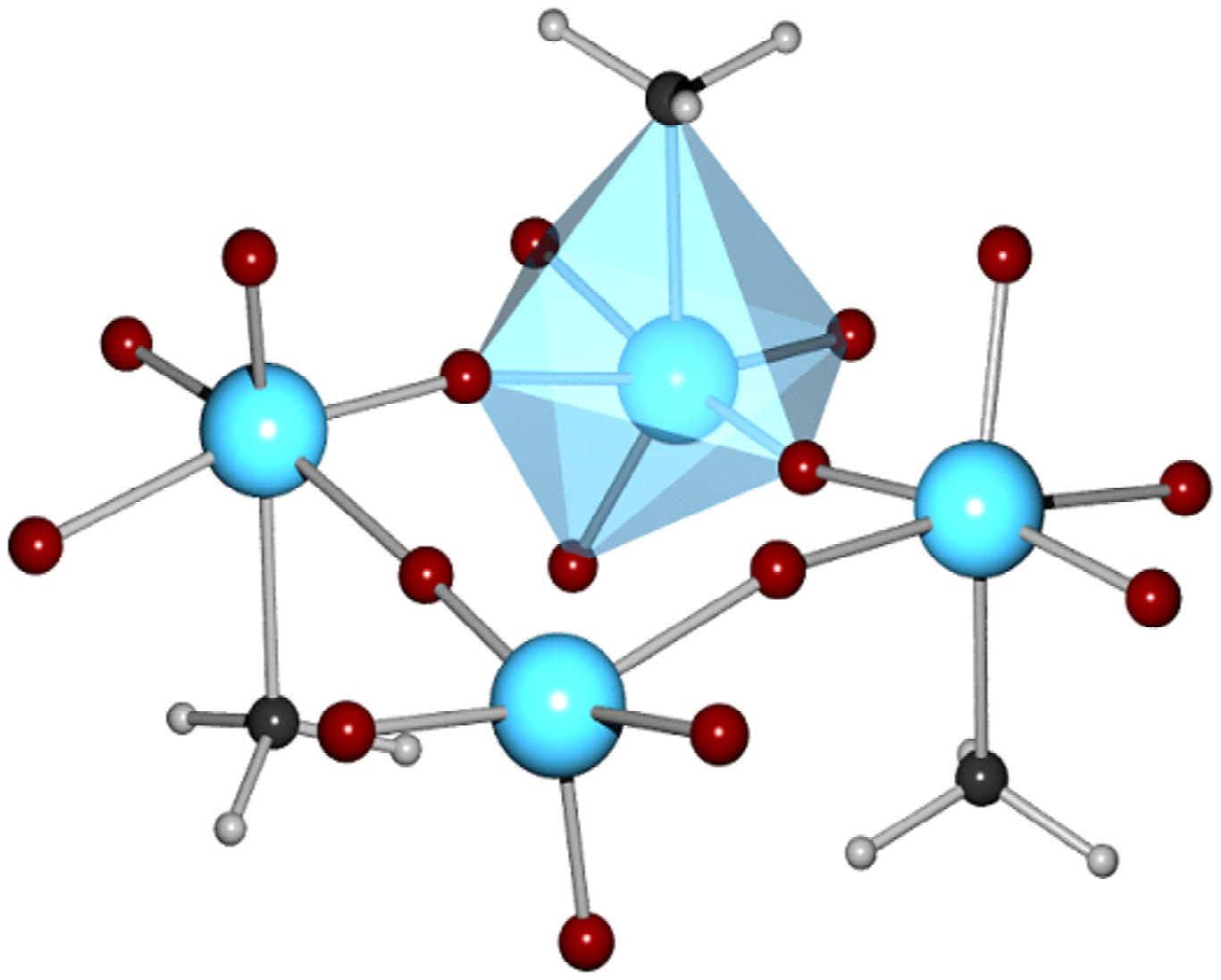}                  
\end{minipage}
\caption{Left: The IR spectrum of \emph{poly}-MTO exhibits an
asymmetric (910\,cm$^{-1}$) and a symmetric (940\,cm$^{-1}$) Re-O
stretching mode which clearly deviates from the corresponding values
in the monomer MTO ($\nu_{as}$(ReO$_3$): 955~cm$^{-1}$;
$\nu_{s}$(ReO$_3$): 1000~cm$^{-1}$). Right: DFT optimized structural
\emph{model~I} of \emph{poly}-MTO (see also Fig.~\ref{fig15}).}
\label{fig3}
\end{center}
\end{figure}

For a closer investigation of the structure of these layers the
geometry of several two-dimensional models of \emph{poly}-MTO were
fully optimized using density functional theory (DFT) methods with
periodic boundary conditions as implemented in the
\textsc{GAUSSIAN03} program-package. \cite{Gaus:04} The PBEPBE
functional in combination with the CRENBL basis-set and an averaged
relativistic effective core potential for Re and a standard 3-21G*
basis-set for C, H and O was used
throughout.\cite{BinkleyRoss:80,Perdew:96} Auxiliary density fitting
functions were generated employing the standard algorithm
implemented in \textsc{GAUSSIAN03}. \cite{Gaus:04} The translational
asymmetric unit was in each case set up as
[\{(CH$_{3}$)ReO$_{3}$\}$_3$ReO$_3$]$_{\infty}$ to account for the
deficiency of methyl groups. However, this leads to a ratio C:Re of
0.75 in our model which is somewhat below the experimental value of
about 0.92 found by the elementary analysis. The models were tested
for different orientation patterns of the methyl groups, the one
leading to the lowest total energy, labeled \emph{model~I}, is shown
in the right panel of Fig.~\ref{fig3}. It should be mentioned that
this model adopts the layered structure of the transition metal
oxide tungstite, WO$_3 \cdot$\,H$_2$O, when replacing the methyl
ligands by water molecules.\cite{Szymanski:84} We note, that the
close structural analogy between tungstite and \emph{poly}-MTO could
be used to employ both compounds as sol-gel precursors for the
synthesis of mixed Re$_{1-x}$W$_x$O$_3$ ceramics by \emph{chimie
douce} methods.\cite{Helbig:05}

The calculated Re--Re distance of 3.79\,\AA\ (average value) is in
good agreement with the distance of 3.67\,\AA\ found in the x-ray
diffraction pattern. Starting from an approximately octahedral
coordination environment of the Re atoms a significant distortion of
the idealized geometry is found. This is indicated by average
O--Re--O angles of 169.3$^{\circ}$ and 154.9$^{\circ}$ in the
\{ReO$_2$\}$_{\infty}$ plane for the CH$_3$ReO$_3$ and the ReO$_3$
fragment, respectively. The  C--Re--O angle clearly deviates from
the ideal angle of 180$^{\circ}$ in the octahedral case and is
reduced to 149.7$^{\circ}$. In addition to the non-planarity of the
\{ReO$_2$\}$_{\infty}$ layer, there also exist two different Re--O
bond distances (1.82\,\AA\ and 2.05\,\AA, average values) for the
methylated rhenium atoms. Each of these Re atoms thus exhibits two
short and two longer Re--O bonds inside the \{ReO$_2$\}$_{\infty}$
plane. This effect is somewhat less pronounced for the ReO$_3$
moiety, for which Re--O bond distances of 1.81\,\AA\ and 1.96\,\AA\
inside the \{ReO$_2$\}$_{\infty}$ plane are found.

\subsection{IR-Spectroscopy}

Infrared-spectra were recorded at room temperature with a Bruker
Fourier-transform spectrometer IFS66v/S. Transmission spectra of
the samples pressed into KBr pellets were collected in the
mid-infrared range from 500 to 5000~cm$^{-1}$. The left panel of
Fig.~\ref{fig3} shows the dominant vibrational double mode at 910
and 940~cm$^{-1}$. Most likely, these two bands arise from the two
Re--O stretching modes (asymmetric (as) and symmetric (s)), which
were found for solid state MTO at 959~cm$^{-1}$
($\nu_{as}$(ReO$_3$)) and 998~cm$^{-1}$ ($\nu_{s}$(ReO$_3$))
prepared in CsI pellets.\cite{mink:94} This is in fine agreement
with room temperature KBr pellet spectra of our MTO precursor
material, which display these two modes at 955 and
1000\,cm$^{-1}$, respectively.

The shift of the vibrational excitations to lower energies for the
polymerized material measured under the same conditions indicates
a weakening of the Re--O bonding in the terminal Re=O groups due
to trans-influence of the CH$_3$ group. This is in line with the
2D structural DFT model discussed above (see Fig.~\ref{fig3})
which shows an elongation of the Re=O bond in \emph{poly}-MTO by
about 0.024\,\AA\ relative to CH$_3$ReO$_3$.

\section{\label{sec:level3}Experimental Results}

\subsection{\label{sec:res}Resistivity}

Resistance measurements were performed by a four-point
low-frequency ac-method below 300\,K over four temperature
decades. For the determination of the resistivity of
\emph{poly}-MTO one has to take into account a rough estimate of
the non-uniform sample shape causing a relative error bar of
$30\%$ for the residual resistivity.

\begin{figure}[t]
\begin{center}
\includegraphics[width=0.48\textwidth]{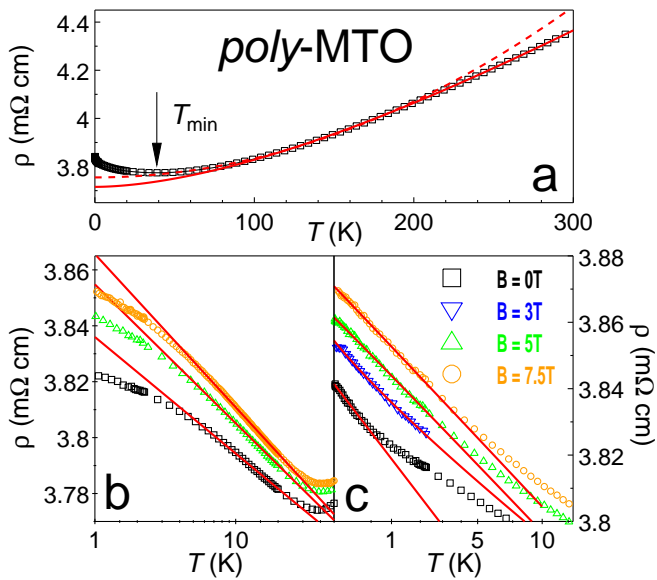}
\caption{a) Resistivity $\rho (T)$ of \emph{poly}-MTO on a linear
scale below 300\,K. The dashed line presents a fit according to
$\Delta\rho (T) \propto T^2$ for $100\,\mathrm{K} > T >
200\,\mathrm{K}$, the solid line is a fit $\Delta\rho (T) \propto
T^2 \ln(T_{\rm{F}}/T)$ for $T>90\,$K (see text). b) A
semi-logarithmic plot of $\rho (T)$ in various magnetic fields.
The solid lines are logarithmic fits between 5\,K and 30\,K. c)
Resistivity $\rho (T)$ plotted vs. square-root $T$ below 10\,K.}
\label{fig4}
\end{center}
\end{figure}

In Fig.~\ref{fig4}a the resistivity of \emph{poly}-MTO is plotted
vs. temperature.  The temperature dependence of the resistivity
clearly exhibits a minimum at about $T_{\rm{min}} = 38$\,K. Above
38\,K the resistivity increases monotonically and non-linearly
with increasing temperature indicating metallic behavior. In order
to identify the dominating scattering mechanisms we analyzed the
data in terms of a power law.\cite{alternative} A least squares
fit according to $\rho(T) = \rho_{0} + A_{\mathrm{ee}} T^2$
between 100\,K and 200\,K (dashed line in Fig.~\ref{fig4}a)
reveals a residual resistivity $\rho_0 = 3.76\,$m$\Omega\,$cm and
an electron--electron scattering amplitude $A_{\mathrm{ee}} = 7.9
\times 10^{-3} \mu \Omega\,$cm\,K$^{-2}$. However, above 200\,K
and below 80\,K the fit deviates substantially from the data.

The absolute value of $A_{\mathrm{ee}}$ is enhanced by a factor of
$10^6$ compared with the value of simple metals,\cite{Kaveh:84}
but is of the same order as found for high-$T_{\rm{c}}$
superconductors.\cite{Tsuei:89} This strong enhancement of the
electron--electron scattering amplitude is most likely due to a
significant anisotropy in our sample. If the 2D nature of
\emph{poly}-MTO --- as derived from x-ray measurements
--- is the origin of such a large enhancement of
$A_{\mathrm{ee}}$, then we have to take into account the relaxation
rate ($1/\tau_{\mathrm{ee}}$) for quasiparticles of a 2D electron
system,\cite{Giuliani:82} which can be expressed as:
\begin{equation}
\frac{1}{\tau_{\mathrm{ee}}} \simeq \left(
\frac{T}{T_{\rm{F}}}\right)^2 \ln \left( \frac{T_{\rm{F}}}{T}
\right) \label{eq:one}
\end{equation}
The resistivity can therefore be written as
\begin{equation}
\rho (T) = \rho_0 + K \left( \frac{T}{T_{\rm{F}}}\right)^2 \ln
\left( \frac{T_{\rm{F}}}{T} \right), \label{eq:two}
\end{equation}
where $K$ is a constant. The solid line in Fig.~\ref{fig4}a is a
fit according to Eq.~(\ref{eq:two}) over the whole temperature
range above 90\,K, leading to a residual resistivity $\rho_0 =
3.75\,$m$\Omega\,$cm and a Fermi temperature $T_{\rm {F}} \simeq
2300$\,K. As can be clearly seen, the fit reproduces the data
convincingly, suggesting that the high temperature resistivity of
\emph{poly}-MTO is actually consistent with Fermi-liquid theories
of electron--electron scattering in pure 2D
metals.\cite{Giuliani:82,Zheng:96} The relatively low value of the
Fermi temperature may be due to the low charge-carrier
concentration and is comparable to values observed for
high-$T_{\rm{c}}$ superconductors (\emph{e.g.} for the classical
hole superconductor YBa$_{2}$Cu${}_3$O$_{7-\delta}$
(Ref.~\onlinecite{Martin:89}) or for the electron-doped
superconductor Nd$_{1.85}$Ce$_{0.15}$CuO$_4$
(Ref.~\onlinecite{Tsuei:89})). The non-linear temperature
dependence of $\rho$ excludes a significant amount of
electron--phonon interaction. From the latter one would expect a
linear temperature dependence above a characteristic temperature
$T^* \approx 0.3\,\Theta_{\mathrm{D}}$; for \emph{poly}-MTO
$\Theta_\mathrm{D}(2\mathrm{D}) = 206$\,K and
$\Theta_\mathrm{D}(3\mathrm{D}) = 66$\,K (see Ref.
\onlinecite{Miller:05}).

\begin{figure}[t]
\begin{center}
\includegraphics[width=0.44\textwidth]{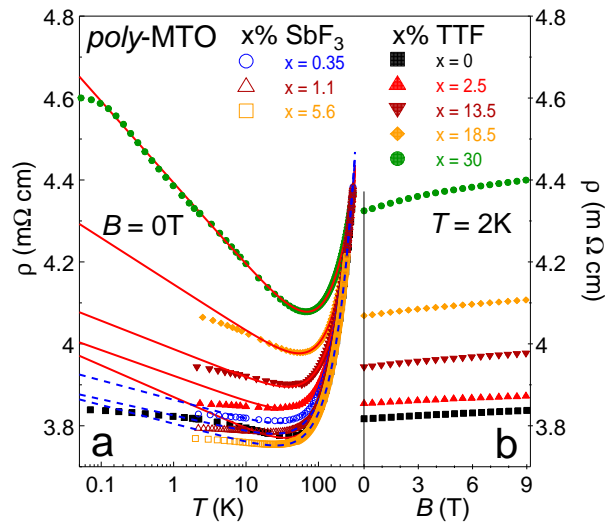}
\caption{a) $\rho$($T$) of \emph{poly}-MTO intercalated with the
organic donor species (TTF) and the inorganic acceptor SbF$_3$.
Notice that the resistivity curves of all samples are normalized
to $\rho$(300\,K) of \emph{poly}-MTO since the complex shape of
the samples prevents a geometrical modelling of their morphology.
The solid (TTF) and dashed (SbF$_3$) lines are fits to the data,
discussed in the text. b) The magnetoresistance of the TTF
intercalated samples at $T = 2$\,K.} \label{fig5}
\end{center}
\end{figure}

Figures~\ref{fig4}b and \ref{fig4}c illustrate the low temperature
resistivity in more details. Below 30\,K a ln$(1/T$) divergence
over one decade of $T$ is observed (Fig.~\ref{fig4}b), and at
lowest temperatures a power--law dependence is detected, $\Delta
\rho \propto T^{\alpha}$, with $\alpha \approx 0.5 \pm 0.1$
(Fig.~\ref{fig4}c). It should be mentioned that with increasing
magnetic field the temperature range of the square root behavior
is expanded. In a recent paper\cite{Scheidt:05} it is suggested
that the origin of the ln$(1/T)$ and $\sqrt{T}$ behaviors may be
due to an additional resistivity contribution related to an
electron–-electron scattering in the presence of a random disorder
potential known as Altshuler-Aronov \cite{Altshuler-Aronov:85}
correction. This will be discussed in more detail in section
\ref{sec:level4}.

In order to increase the concentration of conducting electrons we
performed controlled intercalation of donor and acceptor
specimens. The temperature-dependent electrical resistivity $\rho
(T)$ of \emph{poly}-MTO intercalated with the organic donor
molecule TTF and the inorganic acceptor compound SbF$_3$ is
plotted in Fig.~\ref{fig5}a for various concentrations normalized
to $\rho$(300\,K) of \emph{poly}-MTO.\cite{acceptor} None of the
intercalated compounds exhibits a significant enhancement of the
metallic behavior. The residual resistivity increases with
increasing TTF concentrations, while it is almost constant for the
SbF$_3$ samples. However, all $\rho$($T$) curves clearly exhibit a
crossover from metallic to insulating behavior at
$T_{\textrm{min}}$.

\begin{figure}[t]
\begin{center}
\includegraphics[width=0.44\textwidth]{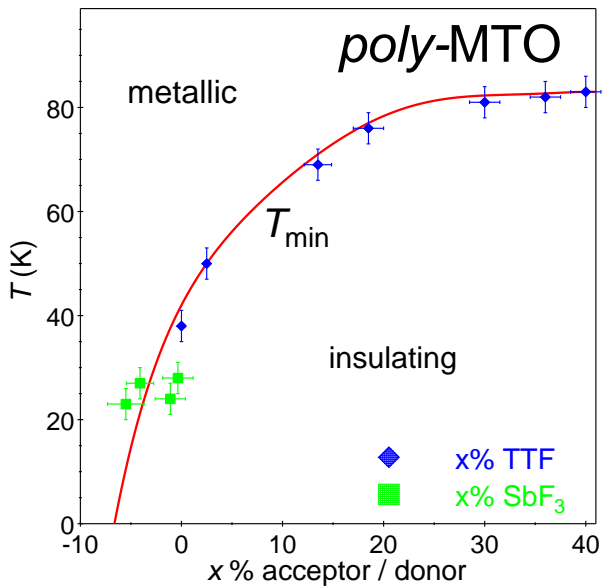}
\caption{Phase diagram  of \emph{poly}-MTO intercalated with $x\%$
of the donor TTF and acceptor SbF$_3$. The line represents the
temperature $T_{\rm{min}}$ determined from the resistivity data in
Fig.~\ref{fig5}.} \label{fig6}
\end{center}
\end{figure}

The solid (TTF) and dashed (SbF$_3$) lines are fits taking into
account Eq.~(\ref{eq:two}) plus an additional ln$(1/T$) term
describing the temperature dependence of $\rho$($T$) below
$T_{\textrm{min}}$. The fits reproduce the data convincingly using a
Fermi temperature of $T_{\rm{F}} = (2500\pm 500)$\,K. It is
conspicuous that for all samples the 2D character of the electron
system is preserved. Furthermore, the logarithmic divergence below
$T_{\textrm{min}}$ becomes more pronounced with increasing TTF
concentration, whereas it remains independent of concentration for
the acceptor specimens doped with SbF$_3$. These different types of
behavior between the donor- and acceptor-intercalated compounds are
in line with the results of the magnetization measurements
indicating that the logarithmic temperature dependence has its
origin in the amount of Re($d^1$) centers (Sec.~\ref{sec:sus_mag},
Tab.~\ref{tab:table1}). For all TTF-intercalated samples a positive,
nearly linearly increasing magnetoresistance was found in this
insulating regime at $T =2$\,K (Fig.~\ref{fig5}b). The same
phenomenon was also observed for pure
\emph{poly}-MTO.\cite{Scheidt:05}

\begin{figure}[t]
\begin{center}
\includegraphics[width=0.44\textwidth]{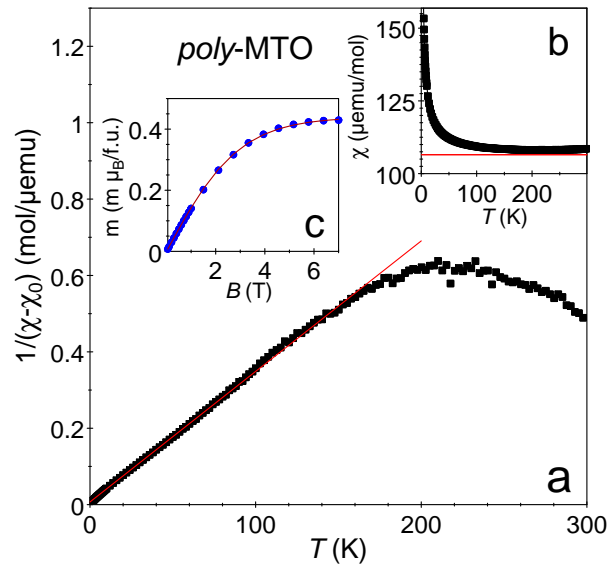}
\caption{a) Inverse magnetic susceptibility of \emph{poly}-MTO
measured in an applied magnetic field of $B = 1$\,T. The solid
line is a fit of the modified Curie-Weiss type behavior. b) The
magnetic \emph{dc}-susceptibility ($\chi =m/B$). c) The magnetic
field dependence of the magnetization. The solid line follows a
Brillouin function for a $d^1$ moment (see text).} \label{fig7}
\end{center}
\end{figure}

A temperature $T_{\textrm{min}}$ was designated as transition
temperature between the metallic (d$\rho$/d$T
>$ 0) and the insulating (d$\rho$/d$T <$ 0) phase (see
Fig.~\ref{fig4}a) to generate a preliminary \emph{phase diagram}
shown in Fig.~\ref{fig6} as a function of TTF and SbF$_3$
concentrations. Here the SbF$_3$ concentration is furnished with a
negative sign to distinguish acceptor character from donor behavior
(positive sign). The concentration dependence of $T_{\textrm{min}}$
of the SbF$_3$ intercalated specimens is rather weak and therefore
SbF$_3$ does not seem to be a suitable compound to follow the
extrapolated phase boundary (solid line in Fig.~\ref{fig6}). This
phase diagram is similar to that found in various
high-$T_{\rm{c}}$-superconductors
\cite{Fournier:98,Boebinger:96,Ono:00} indicating that
\emph{poly}-MTO and its intercalated specimens may be a promising
starting material for the design of superconducting organometallic
polymers.


\subsection{\label{sec:sus_mag}Susceptibility and Magnetization}

Magnetization and susceptibility are excellent measures to get more
information about the partially localized electrons causing a
certain amount of paramagnetic Re$^{\mathrm{VI}}$($d^1$) centers. We
noted before that these electrons as well as the itinerant electrons
in the conduction band originate from the loss of 8$\%$ of the
methyl groups in \emph{poly}-MTO. This section focuses on the role
of these additional electrons originating from this demethylation of
Re and tries to clarify whether these electrons are localized or
itinerant in \emph{poly}-MTO and its intercalated systems. A
superconducting quantum interference device (SQUID) served for the
determination of the magnetization $m$ from 1.9\,K up to 300\,K in
magnetic fields up to 7\,T.

For pure \emph{poly}-MTO the magnetic \emph{dc}-susceptibility
$\chi(T) = m(T)/B$ is pictured in Fig.~\ref{fig7}b. Due to the
small susceptibility values the data are corrected by the
core-diamagnetism $\chi^{\rm{Langevin}}_{\rm{dia}} = - 69 \times
10^{-6}$\,emu/mol.\cite{chi_dia} It is interesting to note that
above 100\,K the susceptibility increases with increasing
temperature, while below 100\,K a paramagnetic contribution
emerges and gets dominant. The unusual high-temperature behavior,
which is present just in the metallic region as revealed by the
resistivity measurements, may be interpreted by a reduction of the
itinerant electron concentration with decreasing temperature.
Another reason could be that the crystal-field splitting of the Re
$d$ orbital manifold is small enough to allow for non-negligible
Van-Vleck contributions to the susceptibility.

Below 70\,K the susceptibility can be well described by a modified
Curie-Weiss type behavior,  $(\chi (T) - \chi_{0}^{\rm{sus}}) =
C/(T-\Theta_{\rm{CW}})$ with a marginal itinerant contribution
$\chi_{0}^{\rm{sus}} = 110 \times 10^{-6}$\,emu/mol, yielding an
average Pauli susceptibility $\chi_{\rm{P}} = 3/2\:
\chi_{0}^{\rm{sus}} = 165 \times 10^{-6}$\,emu/mol (cf.
Fig.~\ref{fig7}a and Tab.~\ref{tab:table1}).\cite{chi_dia}

The vanishing paramagnetic Curie-Weiss temperature $\Theta_{\rm{CW}}
=0$\,K indicates that no correlations between the magnetic moments
of the residual localized electrons at the Re($d^1$) centers are
present. The effective paramagnetic moment $\mu_\mathrm{eff} = 39
\times 10^{-3} \mu_{\rm{B}}$ (Tab.~\ref{tab:table1}) obtained from
the Curie constant $C$ is smaller than expected for
Re$^{\mathrm{VI}}$($d^1$) ions by two orders of magnitude. Assuming
that only integer fractions of Re$^{\mathrm{VI}}$($d^1$) ions exist,
the small effective moment indicates that roughly one
Re$^{\mathrm{VI}}$($d^1$) center per 2000 Re atoms is present. The
inverse of that virtual Re$^{\mathrm{VI}}$ concentration $1/n =
\mu^{2}_{\mathrm{eff[\text{Re}\;d^1]}} /\mu^{2}_{\mathrm{eff[exp]}}$
(with $\mu_{\mathrm{eff[\text{Re}\;d^1]}} = 1.73\,\mu_{\mathrm{B}}$)
is listed in Tab.~\ref{tab:table1}.

In Fig.~\ref{fig7}c the field dependence of the magnetization $m(B)$
after substraction of the core-diamagnetism\cite{chi_dia} and the
itinerant electron contribution ($\chi_{0}^{\rm{mag}}$) at 2\,K is
displayed. The  solid line obeys a Brillouin function taking into
account an Re($d^1$) electron configuration with a quenched orbital
moment ($L = 0$) using $\chi_{0}^{\rm{mag}}$ and the saturation
magnetization ($m_{\rm{s}}$) as fit-parameters. This fit leads to
nearly the same amount of 1/2300 localized Re($d^1$) centers per
formula unit as obtained from the susceptibility measurement (cf.
Tab.~\ref{tab:table1}).

\begin{figure}[t]
\begin{center}
\includegraphics[width=0.44\textwidth]{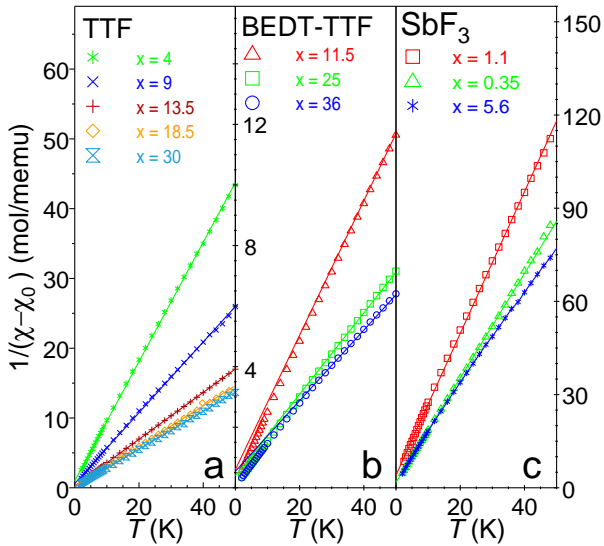}
\caption{Inverse magnetic susceptibility vs. temperature in an
applied field of 1\,T for a) \emph{poly}-MTO$ + x\%\;$TTF, b)
\emph{poly}-MTO$ + x\%\;$BEDT-TTF, and c) \emph{poly}-MTO$ +
x\%\;$SbF$_3$. All lines are fits of the modified Curie-Weiss type
behavior.} \label{fig8}
\end{center}
\end{figure}

\begin{table}[b]
\caption{\label{tab:table1} Localized and itinerant moments
evaluated from susceptibility and magnetization measurements of
\emph{poly}-MTO and its intercalated hybrids with TTF, BEDT-TTF
and SbF$_3$. Note that $\chi_{0}^{\rm{sus}}$ and
$\chi_{0}^{\rm{mag}}$ are given in units of $10^{-6}\;$emu/mol.}
\begin{ruledtabular}
\renewcommand{\arraystretch}{1.3}
\begin{tabular}{lcccccc}
& $x$ & $\chi_{0}^{\rm{sus}}$  / $\chi_{0}^{\rm{mag}}$  & $\mu_{\mathrm{eff[exp]}}$        & $1/n$  & $\mu_\mathrm{B}/m_\mathrm{S}$ \\
& [\%]  &                                               & $[\mu_\mathrm{B}/\mathrm{f.u.}]$ &        &                               \\
\hline
 \emph{poly}-MTO
    &      & 110/120 & 0.039 & 2000 & 2300  \\
 \hline
  TTF
    & 2.5  & 102/126 & 0.065 & 689  & 724   \\
    & 4    &  90/129 & 0.084 & 422  & 440   \\
    & 9    &  92/149 & 0.112 & 240  & 263   \\
    & 13.5 &  72/155 & 0.142 & 148  & 161   \\
    & 18.5 &  71/167 & 0.156 & 125  & 135   \\
    & 30   &  61/164 & 0.161 & 115  & 124   \\
 \hline
BEDT-TTF
    & 11.5 & 68/169  & 0.163 & 112  & 103 \\
    & 25   & 58/145  & 0.206 & 70   & 64  \\
    & 36   & 75/155  & 0.223 & 62   & 62  \\
 \hline
SbF$_3$
    & 0.35 & 120/114 & 0.060 & 833 & 1000  \\
    & 1.1  &  87/110 & 0.049 & 1250& 1600  \\
    & 5.6  &  97/120 & 0.063  &  760&  926  \\
\end{tabular}
\end{ruledtabular}
\end{table}

The inverse susceptibility of TTF-, BEDT-TTF- and
SbF$_3$-intercalated \emph{poly}-MTO samples below 50\,K is
plotted vs. temperature for various concentrations in
Fig.~\ref{fig8}. In the high temperature regime (not displayed)
the weakly intercalated samples follow the same unusual
temperature dependence as \emph{poly}-MTO but in a temperature
range which shifts towards higher temperature with increasing
concentration. The fits of the modified Curie-Weiss law (solid
lines) match the data very well. The resulting values of
$\mu_{\rm{eff[exp]}}$ and $\chi_{0}^{\rm{sus}}$ are listed in
Tab.~\ref{tab:table1}, using the same calculation procedure as
described for pure \emph{poly}-MTO.

\begin{figure}[t]
\begin{center}
\includegraphics[width=0.44\textwidth]{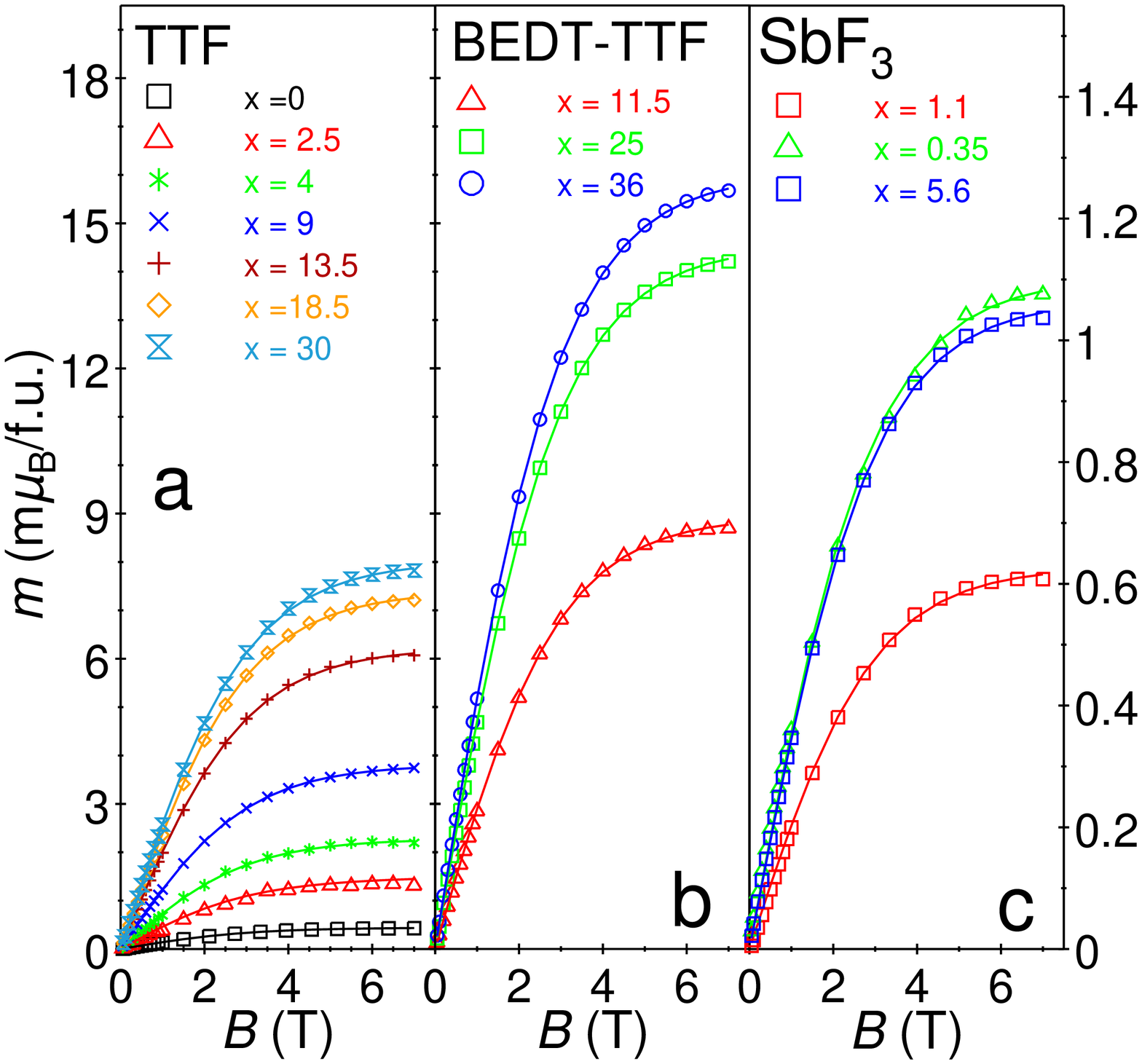}
\caption{Magnetic field dependence of the magnetization at 2\,K
for a) \emph{poly}-MTO$ + x\%\;$TTF, b) \emph{poly}-MTO$ +
x\%\;$BEDT-TTF, and c) \emph{poly}-MTO$ + x\%\;$SbF$_3$. All lines
are fits based on Brillouin functions (see text).} \label{fig9}
\end{center}
\end{figure}

It can be seen from Tab.~\ref{tab:table1} that with increasing donor
concentration the itinerant contribution ($\chi_{0}^{\rm{sus}}$)
decreases while the effective magnetic moment
($\mu_{\mathrm{eff[exp]}}$) increases, mirroring the cumulative
amount of localized electrons as reflected by the presence of
Re$^{\mathrm{VI}}$($d^1$) centers. As a consequence, the number of
Re atoms which share one spin moment ($1/n$), decreases strongly
with increasing $x$ except for SbF$_3$.

This is also supported by the magnetization measurements $m(B)$
displayed in Fig.~\ref{fig9}. The solid lines are fits, following
the Brillouin function for a Re($d^1$) electron configuration,
using the saturation magnetization $m_{\rm{s}}$ and
$\chi_{0}^{\rm{mag}}$ as fit parameters. The enhancement of
$\chi_{0}^{\rm{mag}}$ by a factor of approximately two in
comparison with $\chi_{0}^{\rm{sus}}$ (cf. Tab.~\ref{tab:table1})
may be due to the crossover behavior from a 2D to a 3D system at
about 2\,K as revealed by the resistivity measurements
(Sec.~\ref{sec:res}). For TTF and BEDT-TTF one $\mu_{\rm{B}}$
divided by $m_{\rm{s}}$ (note that $m_{\rm{s}}$ has the unit of
$\mu_{\rm{B}}$) equals roughly $1/n$ derived from the
susceptibility data at high temperatures. Therefore, both
$\chi(T)$ at elevated temperatures and $m_{\rm{s}}$ at 2\,K,
report independently the increasing local character of the
electrons at the Re$^{\mathrm{VI}}$ ions introduced via
intercalation. However, for the SbF$_3$-intercalated samples this
tendency could not be observed within the error bars.

In spite of the increasing carrier localization with increasing
donor concentration no magnetic correlations are observed as
indicated by the vanishing paramagnetic Curie-Weiss temperature
($\Theta_{\rm{CW}} =0$\,K). These results are plausible taking
into account the mean distance
$d_{\rm{Re}^{\mathrm{VI}}-\rm{Re}^{\mathrm{VI}}}$ of the remaining
local Re($d^1$) centers within the planes (e.g. for pure
\emph{poly}-MTO, $d_{\rm{Re}^{\mathrm{VI}}-\rm{Re}^{\mathrm{VI}}}
\approx 140$\,\AA; for $x = 30\%$\:TTF,
$d_{\rm{Re}^{\mathrm{VI}}-\rm{Re}^{\mathrm{VI}}} \approx
40$\,\AA). This large separation of the local magnetic moments is
still too large to form magnetic correlations.

\begin{figure}[t]
\begin{center}
\includegraphics[width=0.48\textwidth]{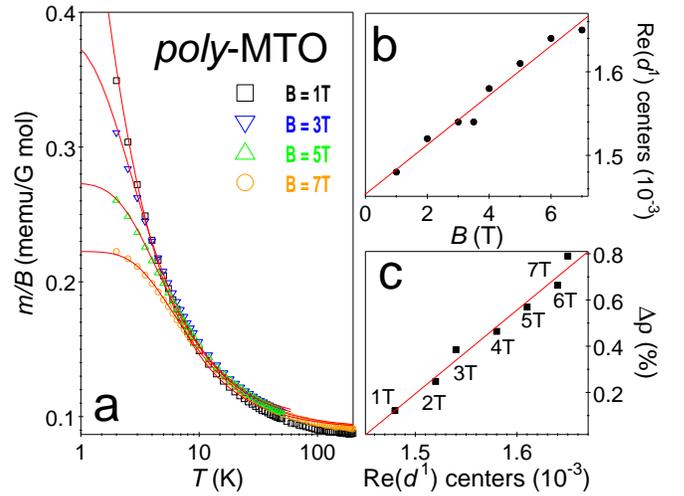}
\caption{a) $m/B$ of \emph{poly}--MTO in several external magnetic
fields. The solid lines are Brillouin-fits for a Re($d^1$)
configuration. b) The amount of Re($d^1$) centers deduced from the
fits vs. $B$. c) Percentage increase of the resistivity with
magnetic field vs. the amount of Re($d^1$) centers.} \label{fig10}
\end{center}
\end{figure}

In a recent paper we suggested that also the magnetic field may
represent a tuning parameter for an additional localization scenario
from which the positive magnetoresistance in pure \emph{poly}-MTO
originates.\cite{Scheidt:05} However, this effect is also observed
in \emph{poly}-MTO samples intercalated by 2.5\,\% TTF. In
Fig.~\ref{fig10}a the magnetization $m$ divided by the applied
magnetic field $B$ is plotted vs. temperature. Again, the solid
lines are Brillouin-fits. The amount of Re($d^1$) centers, formally
deduced from these fits, is pictured in Fig.~\ref{fig10}b. In
Fig.~\ref{fig10}c the relative increase of the magnetoresistance of
\emph{poly}-MTO at 100\,mK is plotted vs. the amount of Re($d^1$)
centers for various fields. Assuming, that the magnetic moment has
its origin only in the amount of the Re($d^1$) centers this linear
behavior is a clear evidence that with increasing magnetic field the
amount of localized electrons and thus the amount of Re($d^1$)
centers increases linearly.

\begin{figure}[t]
\begin{center}
\includegraphics[width=0.44\textwidth]{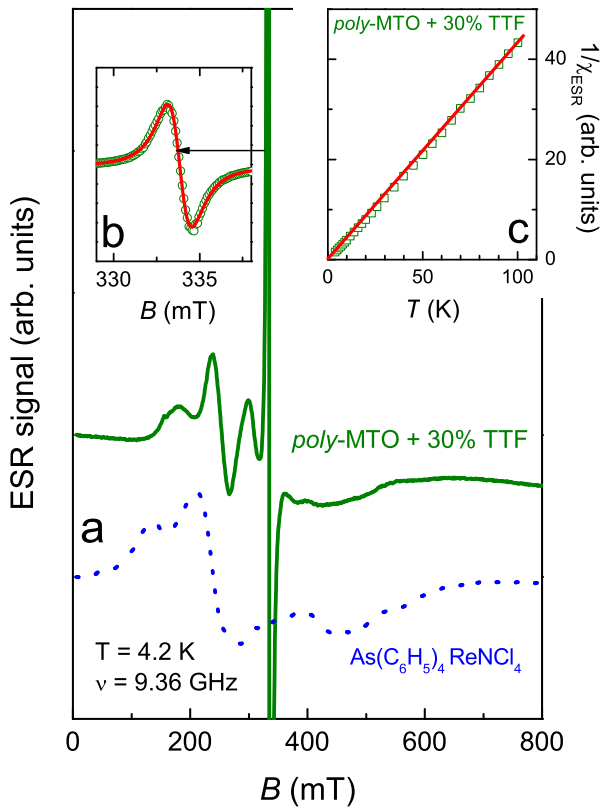}
\caption{a) A typical ESR resonance spectrum of an intercalated
\emph{poly}-MTO sample (solid line). The dotted line exhibits the
spectrum of the reference compound As(C$_6$H$_5$)$_4$ReNCl$_4$. b)
The narrow signal at 0.334\,T corresponds to a $g$-factor of 2.
The solid line is a Lorentz fit to the data. c) The calculated
intensity of the Lorentz curve follows a Curie law (solid line).}
\label{fig11}
\end{center}
\end{figure}

\subsection{ESR-Analysis}

To get more detailed information concerning the origin of the
remaining local electrons, electron spin resonance (ESR)
measurements were performed. The ESR measurements were carried out
at X-band frequencies (9.4\,GHz) with a Bruker ELEXSYS E500-CW
spectrometer using a continuous Helium gas-flow cryostat (Oxford
Instruments) for temperatures $4.2$\,K\:$\leq T \leq 300$\,K. To
avoid the influence of the skin effect due to the
conductivity,\cite{Barnes:81} the polycrystalline samples have been
powdered to grains as small as possible ($< 40\,\mu$m) and immersed
in paraffine. Only for the pure \emph{poly}-MTO sample the grain
size was still not smaller than the skin depth and the influence of
dispersion was visible in the ESR spectra. For all TTF and BEDT-TTF
intercalated compounds, however, the skin effect was negligible.

Electron spin resonance detects the power $P$ absorbed by the sample
from the transverse magnetic microwave field as a function of a
static magnetic field $B$. The signal-to-noise ratio of the spectra
is improved by recording the derivative $dP/dB$ using a lock-in
technique with field modulation. Figure~\ref{fig11}a shows a
spectrum of \emph{poly}-MTO intercalated with 30\% TTF which is
characteristic for all TTF and BEDT-TTF-intercalated samples as
well. All samples exhibit a broad absorption band consisting of
several broad lines with individual linewidths larger than 50\,mT
and a sharp resonance (cf. Fig~\ref{fig11}b) at a resonance field of
0.334\,T corresponding to $g = \hbar\omega/\mu_{\mathrm{B}}B =
2.004$ with a linewidth of approximately 1\,mT. The linewidths and
line positions of the spectra are independent of the temperature,
but their intensities which reflect the corresponding spin
susceptibilities, $\chi_{\rm ESR}$, follow a Curie law as, e.g.,
indicated in Fig.~\ref{fig11}c for the narrow line which can easily
be detected up to high temperatures. This result is in good
agreement with \emph{dc}-susceptibility data (Fig.~\ref{fig8}).

To identify the ESR probes responsible for the observed spectrum we
measured As(C$_6$H$_5$)$_4$ReNCl$_4$ as reference
compound,\cite{Voigt:00} which contains Re in oxidation state +VI
with electronic configuration $5d^1$ as the only magnetic center.
Its ESR spectrum is displayed as dotted line in Fig.~\ref{fig11}a.
One observes a similar broad absorption band with characteristic
lines like those in the \emph{poly}-MTO samples but without the
pronounced sharp resonance at $g = 2$. This broad absorption band
results from the hyperfine interaction of the Re($d^1$) electron
spin $S=1/2$ with the $^{185}$Re, $^{187}$Re nuclear spin $I=5/2$.
In principle the hyperfine structure should consist of six
transitions of equal intensity, but the anisotropy of the hyperfine
constant gives rise to a powder average with some lines more
pronounced than the others.\cite{Gibson:76} From this comparison we
can immediately ascribe the broad absorption band in \emph{poly}-MTO
to $d$ electrons, which are localized in the Re $5d$-shell. The
resolved hyperfine structure corroborates the conclusion from the
Curie law that there are practically no correlations between the
localized electrons at the Re($d^1$) sites.

\begin{figure}[b]
\begin{center}
\includegraphics[width=0.44\textwidth]{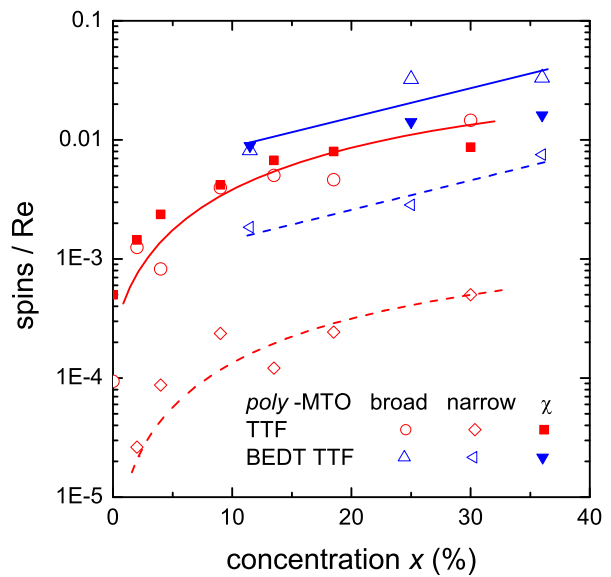}
\caption{The localized moments per Re atom resulting from calculated
intensity of the broad and narrow ESR signals vs. TTF and BEDT-TTF
concentration in comparison with the magnetic susceptibility
findings (Tab.~\ref{tab:table1}). Notice that the narrow band
intensity at $g = 2$ is by a factor of 10 smaller than the broad
band intensity which is attributed to the $5d^1$ moment at the Re
site (all lines are guides to the eye).} \label{fig12}
\end{center}
\end{figure}

At a first glance conduction electrons seem to be accountable for
the observed sharp line. This can be ruled out, since its
intensity follows a Curie law (see Fig.~\ref{fig11}c)
characteristic for localized spins in contrast to a constant Pauli
susceptibility expected for itinerant electrons. In general the
intensity of the narrow line is by a factor of 10 smaller than
that of the broad absorption band (Fig.~\ref{fig12}). The
intercalation dependence of the organic molecules suggests that
these electrons are mainly localized at the TTF and BEDT-TTF
molecules.

The quantitative comparison of the intensities of the spectra in
\emph{poly}-MTO and the reference compound allows to estimate the
number of localized electrons dependent on the degree of
intercalation. In Fig.~\ref{fig12} the resulting local spins per
Re atom are displayed vs. the concentration of TTF and BEDT-TTF.
We find that the intensities increase with TTF as well as with
BEDT-TTF amount. The absolute numbers are in reasonable agreement
with the results obtained from SQUID measurements. For pure
\emph{poly}-MTO the spectra are similar to those of the
intercalated compounds, but due to a marginal amount of localized
electrons they are very noisy. This is  also the case for the
SbF$_3$ intercalated samples. Here the spectra do not show any
dependence on intercalation at all. So far the ESR measurements
allow us to identify Re as the main localization center of the
electrons. This is a convincing evidence for spatial localization
of the electrons at the Re$^{\mathrm{VI}}$($d^1$) centers which is
considered as the origin of the unusual linear positive
magnetoresistance in \emph{poly}-MTO.


\subsection{\label{sec:specheat}Specific heat capacity}

The specific heat measurements on samples with a mass of about
3\,mg were performed at temperatures ranging from 1.8\,K up to
300\,K by means of a quasi-adiabatic step heating technique in
external magnetic fields up to 9\,T using a physical properties
measurement system (PPMS) from Quantum Design. Specific heat data
were also collected at temperatures down to about 80\,mK in a
$^3$He/$^4$He cryostat using a relaxation
method.\cite{Bachmann:72}

\begin{figure}[t]
\begin{center}
\includegraphics[width=0.46\textwidth]{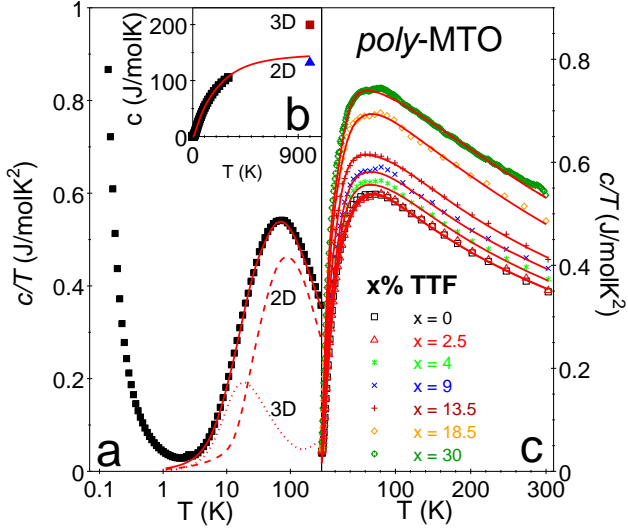}
\caption{ a) Specific heat of \emph{poly}-MTO divided by
temperature vs. $\log T$. The phonon contribution can be well
parameterized using $75 \%$ 2D (dashed line) and $25 \%$ 3D
(dotted line) terms. b) The solid line represents an extrapolation
of the calculated phonon contribution of pure \emph{poly}-MTO. At
1000 K it is closer to the 2D Dulong-Petit limit (triangle) than
to the 3D limit (square). c) $c/T$ vs. $T$ of \emph{poly}-MTO +
$x\%$\,TTF. The solid lines are calculated phonon contributions
and $x$ is the TTF to Re ratio in \%.} \label{fig13}
\end{center}
\end{figure}

Figure~\ref{fig13}a documents the molar specific heat capacity
divided by temperature $c/T$ of \emph{poly}-MTO for temperatures
$0.08\,\mathrm{K} < T < 300$\,K. As the results cover almost four
decades in temperature, for representation purposes, we display
$c/T$ vs. $T$ as semi-logarithmic plot.

Attempts to model the specific heat data by taking into account
only 2D or only 3D Debye and Einstein modes fail. Alternatively,
the phonon contribution can be well parameterized by a sum of 75\%
two- and 25\% three-dimensional terms with
$\Theta_\mathrm{D}(2\mathrm{D}) = 206$\,K and
$\Theta_\mathrm{D}(3\mathrm{D}) = 66$\,K for the low temperature
region (dashed and dotted lines in Fig.~\ref{fig13},
respectively). The high temperature regime is well matched using
additional Einstein terms describing the vibrational modes which
have a predominant molecular character. Especially the highest 3D
Einstein temperature $\Theta_\mathrm{E}(3\mathrm{D}) = 1150$\,K
corresponds well to the IR-active mode at 740\,cm$^{-1}$
associated with the $\rho$-rocking modes of the CH$_3$ groups of
pure MTO.\cite{mink:94,morris:01} In Fig.~\ref{fig13}b the
specific heat $c$ of the calculated phonon contribution of
\emph{poly}-MTO is extrapolated up to 1000\,K in order to compare
the high temperature value with the Dulong-Petit limit of a 2D
(triangle) and a 3D (square) system. Again, the primarily 2D
character of \emph{poly}-MTO is obvious.

\begin{figure}[t]
\begin{center}
\includegraphics[width=0.40\textwidth]{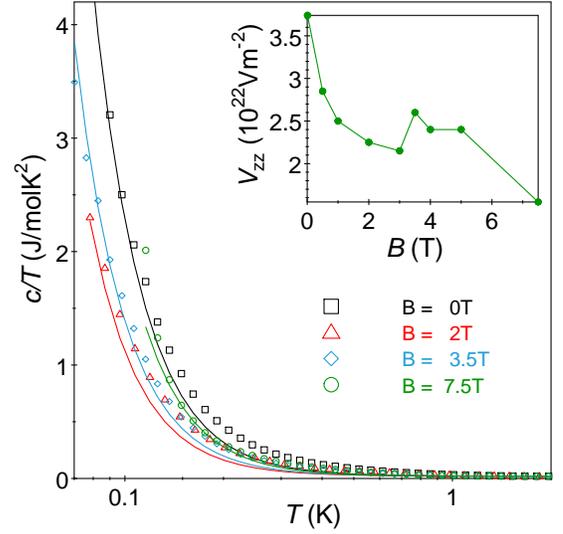}
\caption{ Specific heat divided by temperature vs. log\,$T$ of
\emph{poly}-MTO in various magnetic fields  below 2\,K. The strong
increase of $c/T$ below 1\,K is mainly due to a nuclear magnetic
and quadrupole moment indicating an internal electrical field
gradient $V_{\textrm{zz}}$ at the Re site which decreases with
increasing magnetic field $B$ (inset).} \label{fig14}
\end{center}
\end{figure}

This 2D character of the phonon contribution is also seen for all
samples intercalated with TTF. The  solid lines in
Fig.~\ref{fig13}c display fits of the lattice contribution to the
data using Debye and Einstein terms. Up to $x = 30\%$ the 2D
character increases slightly with decreasing Debye temperatures,
yielding for $30\%$~TTF a 2D to 3D ratio of 4 with
$\Theta_\mathrm{D}(2\mathrm{D}) = 105$\,K and
$\Theta_\mathrm{D}(3\mathrm{D}) = 50$\,K.  Furthermore, the linear
TTF concentration dependence of the high temperature specific heat
$c$(300\,K) clearly demonstrates that TTF is completely
intercalated. These results are in accordance with the fact that
the TTF molecules are positioned in between the \emph{poly}-MTO
layers leading to a slightly expanded layer distance.

The specific heat results of \emph{poly}-MTO + $x \%$~BEDT-TTF
show the same tendency as found for the TTF-intercalated samples.
For example 25\%~BEDT-TTF also yields a 2D to 3D ratio of 4 with
$\Theta_\mathrm{D}(2\mathrm{D}) = 98$\,K and
$\Theta_\mathrm{D}(3\mathrm{D}) = 40$\,K and indicate a slightly
stiffer lattice relative to non-intercalated samples.

The electronic contribution to the specific heat $\Delta c/T$ of
\emph{poly}-MTO was obtained by subtracting the phonon and nuclear
contribution (see below) from the measured $c/T$ value. An
extrapolation of $\Delta c/T$ for $T \to 0$\,K yields $\gamma
\approx (13 \pm 2)$\,mJ\,mol$^{-1}$K$^{-2}$, in agreement with
typical values for $d$-band metals. In addition, the Wilson-ratio
equals $R = \pi^2 k_\mathrm{B}^2/(2 \mu_0 \mu_\mathrm{B}^2) \cdot
\chi_\mathrm{P}/ \gamma \simeq 1$ as it is expected for non-magnetic
metals. The electronic density of states of pure \emph{poly}-MTO
calculated from both, the Pauli-susceptibility $\chi_\mathrm{P} =
1.5 \times \chi_{0}^{\rm{sus}}$ and the Sommerfeld coefficient
$\gamma$, yields $N(E_{\mathrm{F}}) =
5.2$\;states/eV$\,\cdot$\,atom.

In order to analyze the origin of this contribution, electronic
structure calculations based on density functional theory and the
local density approximation (LDA) were performed, which used the
augmented spherical wave (ASW) method.\cite{Williams:79,Eyert:00} As
a preliminary result, a value of $ N(E_{\mathrm{F}}) \approx
1$\;states/eV$\,\cdot$\,atom was obtained.\cite{morecalcs} According
to the calculations, the density of states at the Fermi energy
traces back mainly to Re($5d$) as well as small O($2p$)
contributions. The largest $5d$ contribution stems from the Re site,
which has no CH$_3$ group nearby.

Below 1\,K the specific heat of \emph{poly}-MTO increases
considerably giving rise to a nuclear electronic Schottky effect.
This effect is significantly reduced with increasing TTF
intercalation.\cite{Miller:05} We have calculated the hyperfine
contributions to the heat capacity originating in zero magnetic
field mainly from the quadrupole moments of $^{185}$Re and
$^{187}$Re, and additionally from the nuclear magnetic moments of
$^{1}$H, $^{185}$Re and $^{187}$Re in an applied magnetic field $B$.
The fits are displayed in Fig.~\ref{fig14}, modelling a local
internal electrical field gradient $V_{\mathrm{zz}}$, an average
applied magnetic field and the Zeeman splitting. We note that for
zero magnetic field the enormous increase of $c/T$ is only caused by
the quadrupole moments of $^{185}$Re and $^{187}$Re leading to
$V_{\mathrm{zz}} = (3.7 \pm 0.8) \times 10^{22}$\,Vm$^{-2}$ (inset
of Fig.~\ref{fig14}). The large error of $V_{\mathrm{zz}}$ takes
into account the variance of different samples and the limited
accessible temperature range only including the onset of the
Schottky anomaly. In contrast, the electrical field gradient of the
cubic perovskite ReO$_3$ ($V_{\mathrm{zz}} \simeq 0.18 \times
10^{22}$\,Vm$^{-2}$) is smaller by a factor of 20 than that found
for \emph{poly}-MTO. This result again clearly suggests a high
anisotropy for \emph{poly}-MTO. With increasing magnetic field
$V_{\textrm{zz}}$ decreases and vanishes at around $B = 7$\,T.
Simultaneously, with increasing field the Zeeman splitting of the
nuclear moment of $^{1}$H becomes dominant and leads approximately
to the same enhanced value of $c/T \simeq
2$\,J\,mol$^{-1}$\,K$^{-2}$ at 100\,mK as determined only from
quadrupole moments of $^{185}$Re and $^{187}$Re in zero field.

\begin{figure}[t]
\begin{center}
\includegraphics[width=0.44\textwidth]{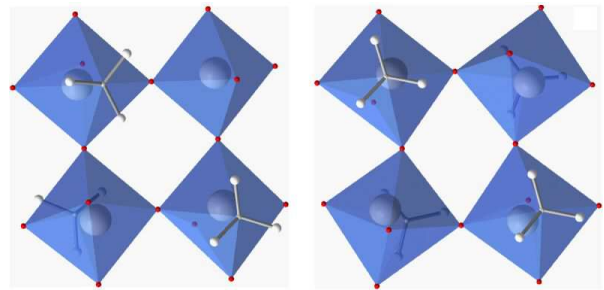}
\caption{Polyhedron representation of the two-dimensional
\emph{models I} (left) and \emph{II} (right) for \emph{poly}-MTO
(view along the $z$-axis perpendicular to the
\{ReO$_2$\}$_{\infty}$ planes) as obtained by DFT calculations.}
\label{fig15}
\end{center}
\end{figure}

The decreasing $V_{\textrm{zz}}$ indicates a reduction of the
electronic and structural anisotropy at the Re sites. DFT
calculations on \emph{poly}-MTO reveal that introducing an
additional electron by demethylation  of $25 \%$ of the Re atoms
leads to a reduction of the asymmetry of the bridging Re--O bonds. A
comparison of the optimized geometries of poly-MTO with one missing
methyl group (\emph{model\;I}; Sec.~\ref{DFT}) and with all methyl
groups present (\emph{model\,II}) is depicted in Fig.~\ref{fig15}.
As a result of the presence of formal Re$^{\mathrm{VI}}$ centers the
difference in the bond length of the bridging Re--O--Re unit inside
the  \{ReO$_2$\}$_{\infty}$ layers is reduced from approximately
0.4\,\AA\ (averaged value) in \emph{model\;II} to approximately
0.2\,\AA\ (averaged value) in \emph{model\;I}. Furthermore,
population analysis of the spin density reveals that the additional
electron which was released by the demethylation process is equally
distributed among the four Re atoms.\cite{Eickerling:05} Thus the
formal chemical reduction
(Re$^{\mathrm{VII}}$\;$\rightarrow$\;Re$^{\mathrm{VI}}$) of one of
the four Re atoms per asymmetric unit causes a clear symmetrization
of the \{ReO$_2$\}$_{\infty}$ planes in \emph{poly}-MTO.
Magnetization and ESR measurements of intercalated samples reveal
that chemical reduction of Re also leads to an increase of electron
localization and thus a larger amount of Re($d^1$) sites. Therefore,
the decrease of $V_{\textrm{zz}}$ with an increasing applied
magnetic field might be directly related to a magnetic field induced
increase of electron localization.


\section{\label{sec:level4}Discussion}

Now we draw our attention to the unusual temperature behavior of
the resistivity of \emph{poly}-MTO below the characteristic
temperature $T_{\mathrm{min}} \simeq 38$\,K (Fig.~\ref{fig4}). We
first focus on the logarithmic increase as it is also observed in
\{CuO$_2$\}$_{\infty}$ layers of some classic cuprates known from
literature (see Sec.~\ref{sec:level1}) and discuss possible
microscopic explanations. Secondly, we will discuss the crossover
(co) at low temperature from the logarithmic temperature
dependence into the square-root dependence of the resistivity at
$T_{\mathrm{co}} \simeq 1.5$\,K .

\begin{table*}[t]
\caption{\label{tab:table2} In-plane resistivities and diffusion
constants for \emph{poly}-MTO and
La$_{1.85}$Sr$_{0.15}$Zn$_{y}$Cu$_{1-y}$O$_4$}
\begin{ruledtabular}
\renewcommand{\arraystretch}{1.3}
\begin{tabular}{lD{.}{.}{2.2}D{.}{.}{2.2}D{.}{.}{1.2}D{.}{.}{2.1}D{.}{.}
{1.1}D{.}{}{3.0}D{.}{.}{1.1}}
 & \multicolumn{1}{c}{$x$;$y$}                      &
\multicolumn{1}{c}{$D_{||}$}               &
\multicolumn{1}{c}{$\rho_{||}$}
 & \multicolumn{1}{c}{$D_\perp$($T_{\textrm{co}}$)} &
\multicolumn{1}{c}{$D_\perp$(3D)}         &
\multicolumn{1}{c}{$L_{||}(T=5$\,K)}
 & \multicolumn{1}{c}{$T_{\textrm{co}}$}\\
 & \multicolumn{1}{c}{[\%]}                         &
\multicolumn{1}{c}{[$10^{-4}\;$m${^2}$/s]} &
\multicolumn{1}{c}{m$\Omega$cm}
 & \multicolumn{1}{c}{ [$10^{-8}\;$m${^2}$/s] }     &
\multicolumn{1}{c}{[$10^{-8}\;$m${^2}$/s]} &
\multicolumn{1}{c}{[\AA]}
 & \multicolumn{1}{c}{[K]}\\
\hline
 \emph{poly}-MTO
     &         & 0.50  & 0.25  & 11   & 2.4    & 87   & 1.5\\
 \hline
  \emph{poly}-MTO + $x$ $\cdot$ TTF
     & 2.5     & 0.50 \footnotemark[1] & 0.25 \footnotemark[1] &       &
& 87. \footnotemark[1]   & \\
     & 13.5     & 0.50 \footnotemark[1] & 0.25 \footnotemark[1] &
&         & 87. \footnotemark[1]  & \\
     & 18.5    & 0.38  & 0.32  &       &         & 76   & \\
     & 30     & 0.24  & 0.51  &  1.5  \footnotemark[2] &         & 60
& 0.2\\
 \hline
 LSCO + $y$ $\cdot$ Zn
    & 0.08  & 11.9   & 0.45 \footnotemark[3]  & 23   &  3.4   & 426   &1 \\
    & 0.10  & 10.52  & 0.61 \footnotemark[3] & 20    & 3.5    & 400   & 0.9 \\
    & 0.12  & 5.9    &  1.27 \footnotemark[3]& 4.6   &  1.0   & 300   & 0.2\\
\end{tabular}
\footnotetext[1]{Within the error bars the gradient of the
logarithmic correction is the same as for \emph{poly}-MTO.}
\footnotetext[2]{For the calculation we used the distance between
the \{ReO$_2$\}$_{\infty}$ layers of \emph{poly}-MTO as  an upper
limit.} \footnotetext[3]{Here we set $\rho_{||}$ equal to
$\rho_{||}$($T_\textrm{min}$) (see Eq.~\ref{LSCO}).}
\end{ruledtabular}
\end{table*}

Two-dimensional weak localization as possible origin of the
$\ln(1/T)$ behavior can clearly be ruled out, due to the magnetic
field dependence of the resistance. A Kondo effect originated by
dilute magnetic impurities might provide a reasonable explanation at
the first glance, because of the small amount of the magnetic
Re$^{\mathrm{VI}}$($d^1$) centers. On the other hand, magnetic
fields should suppress the Kondo effect, whereas experimentally a
positive magnetoresistance is observed up to $B = 7$\,T (see
Figs.~\ref{fig5}b and \ref{fig10}c). Another possible explanation
might arise from a more exotic type of the Kondo effect: a high
temperature $\ln(1/T)$, followed by a low temperature $\sqrt{T}$
behavior, as observed in our experiments, is predicted for systems
displaying two-channel Kondo impurities.\cite{COX:98} The crossover
temperature between $\ln(1/T)$ and $\sqrt{T}$ is expected to be
specific to the microscopic realization of the Kondo impurities.
However, this assumption  is in clear conflict with the fact that
high TTF intercalation degrees strongly reduce the crossover
temperature (see Fig.~\ref{fig5}a). A logarithmic temperature
dependence in the resistivity might also arise in closely related
scenarios as a result of combined disorder and interaction effect
(Altshuler-Aronov corrections):\cite{Altshuler-Aronov:85} (i) the
correction to the diffusive transport in two dimensions; (ii) the
correction to the tunneling conductivity in two dimensions, and
(iii) the correction to the conductivity in systems where the charge
diffusion is effectively \emph{zero
dimensional};\cite{Schwab:02,Schwab:03,Beloborodov:04} the latter
two corrections apply to granular systems.

For \emph{poly}-MTO a precise comparison between theory and
experiment is difficult, since we have no information on the
granularity or possible current paths through the samples. In the
following we discuss in more detail possibility (i) including a
crossover from 2D to 3D diffusion. In a simple ansatz we assume
that
\begin{equation} \label{eq_rho_t}
\rho(T) \approx  \rho_0 + \alpha \rho_{||}(T).
\end{equation}
Here $\rho_0$ represents a temperature independent contribution to
the resistivity from all mechanisms but the resistivity inside the
\{ReO$_2$\}$_{\infty}$ layers. $\rho_{|| } $ is the resistivity
parallel to the \{ReO$_2$\}$_{\infty}$ layers. Due to the
two-dimensional character of \emph{poly}-MTO it is reasonable to
assume that the resistivity parallel to the planes is much smaller
than perpendicular to the planes, $\rho_\perp$. In a macroscopic
sample with arbitrary orientations of grains, the contribution to
the resistance due to diffusion of the electrons parallel to
\{ReO$_2$\}$_{\infty}$ layers is then larger than $\rho_{||}$ but
of the same order. This is encoded in the parameter $\alpha$.
$\alpha$ is larger than one, but still of order of one and for
simplicity we assume that $\alpha \approx 2$. A quantitative
estimation of the resistivity is presented in the following. For
the sample shown in Fig.~\ref{fig4}b the temperature dependence of
the resistivity in the temperature range between 2\,K and 30\,K
and in zero magnetic field obeys
\begin{equation}
\rho(T) = [ 3.75 + 2\times 10^{-2} \ln( T_{\textrm{min}} / T) ] \,
\, {\rm m\Omega\,cm},  \label{2d_experiment}
\end{equation}
with $T_{\textrm{min}} = 38$\,K. This has to be compared with the
theoretical estimate
\begin{equation}
\rho(T) \approx \rho_0 + \frac{\alpha}{e^2 D_{\mathrm{
||}}N(E_{\mathrm{F}})+ \delta \sigma_{||}(T)}, \label{rho(T)_2d}
\end{equation}
where $D_{||}$ is the electron diffusion constant parallel to the
\{ReO$_2$\}$_{\infty}$ layers, and $N(E_{\mathrm{F}})$ is the
density of states at the Fermi energy, given by $N(E_{\mathrm{F}})
= 5.2$\;states/eV$\,\cdot$\,atom $= 5.1 \times
10^{28}$\,states/eV$\,\cdot$\,m$^3$ (see Sec.~\ref{sec:specheat}).
An expansion of Eq.~(\ref{rho(T)_2d}) for small $\delta
\sigma_{||}(T)$ results in
\begin{equation}
\rho(T) \approx \rho_0 + \alpha \rho_{||}(T_{\mathrm{min}})-
\frac{\alpha}{(e^2 D_{\rm ||}N(E_{\mathrm{F}}))^2} \cdot \delta
\sigma_{||}(T). \label{2d_entwickelt}
\end{equation}
In two-dimensional systems the Altshuler-Aronov
correction\cite{Altshuler-Aronov:85} (disorder enhanced
electron--electron scattering) to the conductivity is
\begin{equation}
\delta \sigma_{||}(T) = - \frac{e^2}{2 \pi^2 \hbar} \frac{1}{d}
\ln(T_{\textrm{min}}/T), \label{Altshuler_corr_2D}
\end{equation}
where $d= (7.4\pm 0.4)$\,{\AA} is the distance between the
\{ReO$_2$\}$_{\infty}$ layers.\cite{distance} Inserting
Eq.~(\ref{Altshuler_corr_2D}) into Eq.~(\ref{2d_entwickelt}) and
comparing the theoretical temperature dependence of the
resistivity with the experimental finding
(Eq.~(\ref{2d_experiment})) enables us to estimate $D_{||} \approx
5 \times 10^{-5}$ m$^2$/s (with $\alpha = 1 $ we obtain $D_{||}
\approx 3.5 \times 10^{-5} $ m$^2$/s) and $\rho_{||} \approx 0.25
\, \, {\rm m \Omega\,cm} $, i.e., six percent of the total
resistance arises due to diffusion in the planes. From the
diffusion constant we determine the thermal diffusion length
$L_{||}= \sqrt{\hbar D_{||}/k_{\mathrm{B}}T} = 87$\,\AA\; at 5\,K
(see Tab.~\ref{tab:table2}) which is clearly smaller than the size
of the grains seen in the micrograph in Fig.~\ref{fig1}.

Upon lowering the temperature the diffusion becomes
three-dimensional with the crossover temperature $k_B
T_{\mathrm{co}} \approx \hbar D_\perp /d^2 $, where $D_\perp$ is
the diffusion constant perpendicular to the planes. Experimentally
the crossover between a high temperature $\ln (1/T)$ and a low
temperature $\sqrt{T}$ behavior in the resistivity is observed
near 1.5\,K, from which we obtain $D_\perp \approx 11 \times
10^{-8}$ m$^2$/s, and which is considerably smaller than $D_{||}$
(see Tab.~\ref{tab:table2}). An independent estimate of $D_\perp$
is obtained from the amplitude of the low temperature square root
dependence of the conductivity ($\delta \sigma_{||}(T)$). Theory
predicts for a system with anisotropic three-dimensional diffusion
\begin{equation}
\delta \sigma_{||}(T) = 0.915 \frac{e^2}{2 \pi^2 \hbar}
\frac{2}{3} \frac{D_{||}}{\overline{ D } } \left( \frac{k_B
T}{\hbar \overline{ D }} \right)^{1/2},
\label{delta_sigma_parallel}
\end{equation}
with an average diffusion constant $\overline{D}= ( D_{||}^2
D_\perp)^{1/3}$, cf. Ref.~\onlinecite{Altshuler-Aronov:85} and
\onlinecite{Bhatt:85}. The experimental value is $\delta \rho( T) =
2.2 \times 10^{-7}\, \, \Omega {\rm  m } \sqrt{T/{\rm K }}$, leading
to $D_\perp = 2.4 \times 10^{-8}$ m$^2$/s ($\alpha = 1$: $D_\perp =
0.6 \times 10^{-8}$ m$^2$/s) using Eq.~(\ref{2d_entwickelt}). For
comparison the estimated values of the diffusion constant $D_\perp$
as deduced in two different ways are listed in
Tab.~\ref{tab:table2}: (i) $D_\perp$($T_{\textrm{co}}$) from the
crossover temperature $T_{\mathrm{co}}$ between the $\ln (1/T)$ and
$\sqrt{T}$ behavior and (ii) $D_\perp$(3D) from the coefficient of
the $\sqrt{T}$ behavior in Fig.~\ref{fig4}c. Considering the crude
nature of Eq.~(\ref{eq_rho_t}), we believe that these results agree
surprisingly well.

\emph{Poly}-MTO intercalated with TTF are treated in the same way
using the data presented in Fig.~\ref{fig5}a. The results are
listed in Tab.~\ref{tab:table2} as well. While $\rho_{||}$ is
increasing, $D_{||}$ is decreasing with increasing TTF amount,
attributed to an increasing number of localized electrons at the
Re($d^1$) sites, which is in line with our scenario. Furthermore,
the crossover temperature $T_{\rm{co}}$ is decreasing with
increasing amount of scattering centers which is reasonable if the
distance between the layers increases with increasing TTF
concentration ($k_B T_{\mathrm{co}} \approx \hbar D_\perp /d^2 $).

\begin{figure}[t]
\begin{center}
\includegraphics[width=0.44\textwidth]{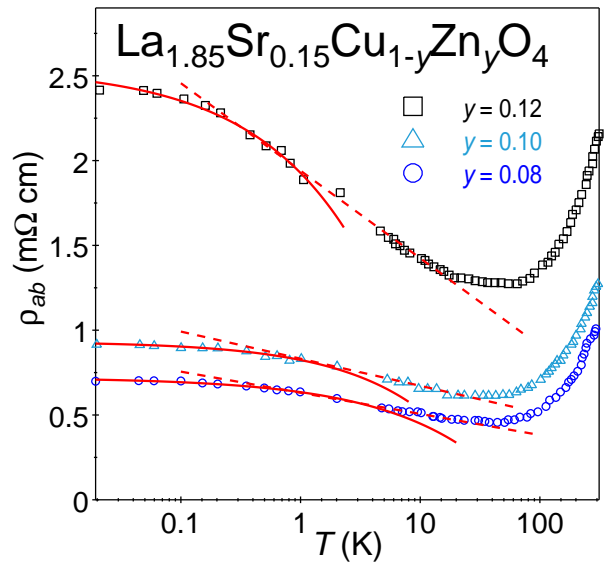}
\caption{The temperature dependence of the resistivity of
La$_{1.85}$Sr$_{0.15}$Cu$_{1-y}$Zn$_y$O$_4$ with $y =
0.08,0.1,0.12$. The data are taken from
Ref.~\onlinecite{Karpinska:00}. The lines are fits to the data
indicating disorder enhanced electron--electron scattering as
discussed in the text (solid lines: $\rho(T) \sim \sqrt{T}$;
dashed lines $\rho(T) \sim \ln(1/T)$).\cite{Altshuler-Aronov:85}}
\label{fig16}
\end{center}
\end{figure}

Finally we comment on the resistivity of
La$_{1.85}$Sr$_{0.15}$CuO$_4$ doped with Zn, since we believe that
the temperature dependence can be explained by the same mechanism
as in {\em poly}-MTO. Experimental data, taken from
Ref.~\onlinecite{Karpinska:00}, are shown in Fig.~\ref{fig16}.
Already in Ref.~\onlinecite{Karpinska:00} it has been noted that
the increase in the resistivity is logarithmic in a certain
temperature range and it was speculated, that the saturation of
the resistivity below 300\,mK could signal some remanence of the
superconducting phase.

Using the microscopic picture from above we start with the ansatz
\begin{equation}
\rho_{||}(T) = \rho_{||}(T_{\rm min}) - \alpha \rho_{||}(T_{\rm
min})^2 \delta \sigma_{||}(T), \label{LSCO}
\end{equation}
and taking $\delta \sigma$ from Eq.~(\ref{Altshuler_corr_2D}) the
logarithm is consistently explained with $\alpha = 2.85$, 2 and
1.45 for the three Zn concentrations $y = 0.08$, 0.1 and 0.12,
respectively (here the distance between the \{CuO$_2$\}$_{\infty}$
planes is $d=13.2$\,\AA).\cite{Day:87}

Figure~\ref{fig16} demonstrates that the logarithmic increase of
$\rho(T)$ changes -- as in {\em poly}-MTO -- to a $\sqrt{T}$
behavior, which we believe is controlled by the crossover from 2D
to 3D electron diffusion. Following the same procedure as before
we estimate the electron diffusion constants parallel and
perpendicular to the \{CuO$_2$\}$_{\infty}$ layers. The results
are listed in Tab.~\ref{tab:table2}. To obtain the diffusion
constant from the conductivity we need the density of states at
the Fermi energy which we assume to be given by $N(E_{\mathrm{F}})
\approx 0.36$\,states/eV$\,\cdot$\,atom $= 9.78 \times
10^{26}$\,states/eV$\,\cdot$\,m$^3$,
Ref.~\onlinecite{Schossmann:87}. The ratio $D_{\perp}/D_{||}$,
which is the ratio of the conductivity perpendicular and parallel
to the \{CuO$_2$\}$_{\infty}$ planes, is thus estimated to be of
the order $10^{-4}$, which is a reasonable result for
La$_{1.85}$Sr$_{0.15}$Zn$_{y}$Cu$_{1-y}$O$_4$.

\section{\label{conclusion}Conclusion}

Summarizing, the inherently conducting organometallic polymer
\{(CH$_{3}$)$_{0.92}$ReO$_{3}$\}$_{\infty}$ (\emph{poly}-MTO) is a
promising prototype for a transition metal oxide featuring
physical properties characteristic for a two-dimensional system.
The 2D character of \emph{poly}-MTO is also reflected by the
unusual diffraction pattern which can be indexed by a square
lattice ($a = 3.67(2)$\,\AA) in conformity with the 2D space group
\emph{p}4\emph{mm}. The pronounced peak shape asymmetry is another
characteristic indicator for a layered compound lacking a 3D
ordering. We therefore suggest a turbostatic or a 00\emph{l}
defect stacking model for the type of disorder, displayed by
\emph{poly}-MTO layers along the crystallographic \emph{c} axis.
The geometry of the corner sharing CH$_3$ReO$_5$ octahedra within
the \emph{poly}-MTO layers was optimized by DFT methods resulting
in a tungstite-type structural model with alternating Re--CH$_3$
methyl groups located above and below the \{ReO$_2$\}$_{\infty}$
layers, respectively. The presence of methyl groups thus hinders
the formation of a 3D oxide network. As a result the layers are
nearly ideally decoupled and solely connected by subtle van der
Waals interactions which is reflected particularly by the large 2D
phonon contribution in the specific heat. This 2D structure is
also confirmed by electrical transport properties. The logarithmic
correction to the temperature dependence of the resistivity at
high temperature is an important manifestation of a 2D Fermion
system.

Below a characteristic temperature $T_{\mathrm{min}}$ the
resistivity changes from metallic to insulating behavior. This
property is intensified by intercalation of the \emph{poly}-MTO
host lattice with donor and acceptor guest molecules leading to a
phase diagram which is related to that found for Zn doped
high-$T_c$ superconductors such as
La$_{1.85}$Sr$_{0.15}$CuO$_{4}$. The origin of metallic behavior
of \emph{poly}-MTO is  mainly due to a Re($5d$) contribution at
the Fermi level as determined by DFT- and LDA-studies. This
preliminary result on the calculated electronic density of states
($ N(E_{\mathrm{F}}) \approx 1$\;states/eV$\,\cdot$\,atom) is
roughly of the same order than that obtained from specific heat
measurements ($N(E_{\mathrm{F}}) \approx
5.2$\;states/eV$\,\cdot$\,atom).

The area of the insulating regime is directly related to the
amount of localized magnetic moments which was analyzed by means
of susceptibility and magnetization measurements. In addition,
ESR-studies have clearly identified that the main part of these
localized electrons is due to the presence of Re($d^1$) centers.
This result is very important to explain the unexpected positive
magnetoresistance in the insulating regime. The susceptibility and
ESR measurements on TTF intercalated \emph{poly}-MTO samples
clearly demonstrate that a stronger formation of the positive
magnetoresistance is related to an increase of spatially localized
electrons. This result is also supported by magnetic field
dependent specific heat studies of the nuclear contribution of the
$^{185}$Re and $^{187}$Re quadrupole moments. The electrical field
gradient $V_{zz}$ in the vicinity of the Re atoms decreases with
increasing external magnetic field indicating a reduction of
distortion. This is corroborated by DFT-calculations which
demonstrate that an enhanced amount of spatially localized
electrons leads to a higher symmetry in the vicinity of the Re
atoms.

Furthermore, we have now a comprehensive understanding of the
logarithmic temperature dependence ln($1/T$) followed by a
square-root behavior of the resistivity at low temperatures. Due
to spin localization a disorder enhanced electron--electron
interaction, as discussed by Altshuler and Aronov, models this
behavior very well taking into account a crossover from a 2D to a
3D system at very low temperatures. Due to the large chemical
variation possibilities, \emph{poly}-MTO and its intercalated
species represent new and promising candidates to stimulate the
field of functional two-dimensional systems.

This work was supported by the Deutsche Forschungsgemeinschaft
(DFG) through the SFB~484.


\begin{thebibliography}{99}
\bibitem{radiation}{For patterns i--iv a Bragg-Brentano device with
Cu$\,K_{\alpha}$ radiation, for pattern v a Guinier camera with
Mo$\,K_{\alpha}$ radiation has been used.}
\bibitem{alternative} In a previous publication \cite{Scheidt:05} the high
temperature resistivity data were fitted with a $T^{3/2}$
dependence. Since no convincing model for this $T$-behavior
exists, we present an alternative approach in this paper.
\bibitem{acceptor} Note the surprising result that
acceptor ligands intercalate a rather electron deficient oxide
host lattice at all!
\end{thebibliography}
\end{document}